\documentclass[aps,prl,preprint,superscriptaddress,nofootinbib]{revtex4-2}
\pdfoutput=1
\usepackage{lipsum}
\usepackage{textgreek}
\usepackage{xcolor}
\usepackage{comment}
\usepackage{graphicx}
\usepackage{subfigure}
\usepackage{mwe}
\usepackage{amsmath}
\usepackage{bm}
\usepackage{mathtools}
\usepackage{braket}
\usepackage{float}
\usepackage[hidelinks]{hyperref}
\usepackage{cleveref}
\usepackage{amsfonts}

\usepackage[export]{adjustbox}
\newcommand\varpm{\mathbin{\vcenter{\hbox{%
  \oalign{\hfil$\scriptstyle+$\hfil\cr
          \noalign{\kern-.3ex}
          $\scriptscriptstyle({-})$\cr}%
}}}}
\newcommand\varmp{\mathbin{\vcenter{\hbox{%
  \oalign{\hfil$\scriptstyle-$\hfil\cr
          \noalign{\kern-.3ex}
          $\scriptscriptstyle({+})$\cr}%
}}}}

\begin{document}


\title{Control of Yu-Shiba-Rusinov States through a Bosonic Mode}


\author{Helene M\"uller}
\email[]{Helene.Mueller@dipc.org}
\thanks{Current address: Donostia International Physics Center (DIPC), Paseo Manuel de Lardiz\'{a}bal 4, 20018 San Sebasti\'{a}n, Spain}
\affiliation{Department of Physics, University of Erlangen-Nuremberg, 91058 Erlangen, Germany}
\affiliation{Max Planck Institute for the Science of Light, Staudtstrasse 2, 91058 Erlangen, Germany}
\author{Martin Eckstein}
\email[]{Martin.Eckstein@fau.de}
\affiliation{Department of Physics, University of Erlangen-Nuremberg, 91058 Erlangen, Germany}
\author{Silvia {Viola Kusminskiy}}
\email[]{S.Viola.Kusminskiy@physik.rwth-aachen.de}
\affiliation{Institute for Theoretical Solid State Physics, RWTH Aachen University, 52074 Aachen, Germany}
\affiliation{Max Planck Institute for the Science of Light, Staudtstrasse 2, 91058 Erlangen, Germany}


\date{\today}
\renewcommand{\abstractname}{\vspace{\baselineskip}}
\begin{abstract}
We investigate the impact of a bosonic degree of freedom on Yu-Shiba-Rusinov (YSR) states emerging from a magnetic impurity in a conventional superconductor. Starting from the Anderson impurity model, we predict that an additional p-wave conduction band channel opens up if a bosonic mode is coupled to the tunnelling between impurity and host, which implies an additional pair of odd-parity YSR states. The bosonic mode can be a vibrational mode or the electromagnetic field in a cavity. The exchange couplings in the two channels depend sensitively on the state of the bosonic mode (ground state, few quanta or classically driven Floquet state), which opens possibilities for phononics or photonics control of such systems, with a rich variety of ground and excited states.
\end{abstract}


\maketitle
 \newpage
A spin-polarised impurity embedded in a s-wave superconductor creates spatially localised and spin-polarised states inside the superconducting gap $\Delta$, the Yu-Shiba-Rusinov (YSR) states \cite{Yu1965,Shiba1968,Rusinov1969,Yazdani1997,Balatsky2006}. Such systems have recently gained renewed attention, since they provide a promising avenue towards realising unconventional superconducting phases and Majorana bound states \cite{Choy2011,Nakosai2013,Perge2013,Nadj-Perge2014,Perrin2020}. This raises the question of how YSR states can be manipulated in a controlled manner \cite{Schneider2021,Akkaravarawong2019,Franke2011,Farinacci2018}. A versatile pathway to control properties of solids is periodic driving (Floquet engineering), which can be used to manipulate both quantum impurities \cite{Nakagawa2015,Iwahori2016,Eckstein2017,Quito2021} as well as bulk properties \cite{Basov2017,Torre2021} such as band structures \cite{McIver2020,Wang2013,Oka2019}, magnetic interactions \cite{Mentink2015,Bukov2015,Claassen2017} and superconductivity \cite{Sentef2017,Babadi2017,Murakami2017,Curtis2022,Buzzi2020}. However, strong driving with time-dependent classical electromagnetic fields inevitably leads to heating via (multi-)photon absorption. A promising alternative direction is to enhance the light-matter coupling \cite{Kockum2019,Diaz2019} by compressing the mode volume in a cavity to the extent that single photons or vacuum fluctuations affect the material properties \cite{Thomas2019,Schlawin2019,Li2020,Curtis2019,Sentef2020,Mazza2019,Sentef2018,Schlawin2022}. By embedding the impurity which gives rise to YSR states into a cavity, the electromagnetic field can either couple directly to the electron tunnelling between the impurity and the host, or through infrared active vibrations which in turn interact with the electronic states at the impurity. For a magnetic impurity in the Anderson model coupled to two normal metal leads, both Floquet engineering by a classical electric field \cite{Eckstein2017,Quito2021} and coupling to a center of mass vibrational mode \cite{Silva2009} have been predicted to give rise to two-channel Kondo physics. One can therefore expect that also strong coupling to a photon or phonon can open additional channels for the emergence of YSR states. This could be used to control multi-channel YSR physics, which can arise in undriven systems due to exchange scattering in different orbital channels \cite{Rusinov1969,Ji2008,Ruby2016,Choi2017,Arrachea2021,Saunderson2022}, due to spin-orbit coupling \cite{Beck2021,Schneider2021} or from two bands in a superconductor \cite{Moca2008}.

In this work, we show that a bosonic mode can activate an additional exchange scattering channel for a magnetic impurity embedded in a superconductor, and investigate its impact on the YSR states. For the coupling in an initial s-wave channel, present for transition metal impurities \cite{Saunderson2022}, the boson-assisted tunnelling involves conduction band states of p-wave symmetry, and therefore corresponds to a distinct conduction band channel in the presence of inversion symmetry. As a consequence, an additional and independent pair of YSR states with p-wave symmetry appears. Our results suggest the opportunity to control the interactions in both channels through a cavity or a classical drive. This paves the way towards phononics or photonics control of such systems, with a tunability of the energy and spin-polarisation of the YSR states as well as the nature of the ground state.

{\bf Model --} We start with a single-orbital Anderson impurity which is embedded in a s-wave superconductor. Assuming inversion symmetry around the impurity, conduction electron states come in pairs related to space inversion, which we refer to as $\pm \mathbf{k}$. Fermion operators $\hat{c}_{\mathbf{k}\sigma}$ are then decomposed into even $\left(\gamma=+\right)$ and odd $\left(\gamma=-\right)$ parity combinations, $\hat{a}_{\mathbf{k}\sigma,\gamma}=\frac{1}{\sqrt{2}}\left(\hat{c}_{\mathbf{k}\sigma} +\gamma \hat{c}_{-\mathbf{k}\sigma} \right)$. In this representation, the Hamiltonian reads
\begin{align}
\hat{H} & = \hat{H}_{d}+\hat{H}_{\text{host}}+\hat{H}_{\text{hyb}}+\hat{H}_{\omega_0}, \label{eq:H}\\
\hat{H}_{d} & = \sum_{\sigma}(\epsilon_{d}-\mu)\hat{n}_{d\sigma}+U \hat{n}_{d\uparrow}\hat{n}_{d\downarrow}, \label{eq:Hd}
\\
\hat{H}_{\text{host}} & = \sum_{\mathbf{k}\sigma\gamma}E_{\mathbf{k}}\hat{\alpha}_{\mathbf{k}\sigma,\gamma}^{\dagger}\hat{\alpha}_{\mathbf{k}\sigma,\gamma}, \label{eq:HSC}
\\
\hat{H}_{\text{hyb}} & = \sqrt{2}\sum_{\mathbf{k}\sigma\gamma}\left[  V_{\mathbf{k},\gamma}(g\hat{Q})\hat{a}_{\mathbf{k}\sigma,\gamma}^{\dagger}\hat{c}_{d\sigma} + h.c.\right]  ,\label{eq:Hhyb} \\
\hat{H}_{\omega_0} & = \omega_0 \hat{b}^\dagger \hat{b}\,\label{eq:Hbos}.
\end{align}
Here, $\hat{H}_{d}$ describes the impurity orbital (with fermion operators $\hat{c}_{d\sigma}^{\dagger}$, $\hat{n}_{d\sigma}=\hat{c}^{\dagger}_{d\sigma}\hat{c}_{d\sigma}$ and spin index $\sigma$) at the energy level $\epsilon_d$ well below the Fermi energy $\mu=0$, and with Coulomb repulsion $U$. For simplicity, we consider the particle-hole symmetric case, with $\epsilon_d=-U/2$. (An asymmetric model would lead to an additional potential scattering term in the low energy Hamiltonian, which plays a subordinate role here, since it implies an asymmetry in the bound state wavefunction but does not create intra-gap states by itself.) The superconductor is included in Eq.~\eqref{eq:H} as a BCS mean-field Hamiltonian $\hat{H}_{\text{host}}$, written in terms of Bogoliubov quasiparticle operators $\hat{\alpha}_{\mathbf{k}\sigma,\gamma}^{\dagger}=u_{\mathbf{k}}\hat{a}_{\mathbf{k}\sigma,\gamma}^{\dagger}-\sigma v_{\mathbf{k}}\hat{a}_{-\mathbf{k}-\sigma,\gamma}$ with excitation energy $E_{\mathbf{k}}=\sqrt{(\epsilon_{\mathbf{k}}-\mu)^{2}+|\Delta|^{2}}$, where $\epsilon_{\mathbf{k}}$ is the single-particle energy and $u_{\mathbf{k}}$ and $v_{\mathbf{k}}$ are the usual coherence factors. The term $\hat{H}_{\text{hyb}}$ describes the tunnelling between the host and the impurity. We consider the coupling to an inversion-odd mode with displacement $\hat{Q}=(\hat{b}^\dagger+\hat{b})/\sqrt{2}$, such as the electric field or vector potential of the cavity, or an infrared active phonon; $\hat{H}_{\omega_0}$ refers to the free Hamiltonian of the bosonic mode $\left(\hbar=1\right)$. The tunnelling matrix element $V_{\mathbf{k}}(g\hat{Q})$ between the impurity orbital and the host states $\mathbf{k}$ depends on $\hat{Q}$, where $g$ is an overall dimensionless coupling strength. Under inversion symmetry $\left[V_{\mathbf{k}}\left(\mathbf{r}\right)=V_{-\mathbf{k}}\left(-\mathbf{r}\right)\right]$, it can be decomposed into the even and odd channel, 
\begin{align}
V_{\mathbf{k},\gamma}(g\hat{Q}) = \frac{1}{2} \left[ V_{\mathbf{k}}(g\hat{Q}) +\gamma V_{\mathbf{k}}(-g\hat{Q}) \right],\label{eq:VCh}
\end{align}
leading to the hybridisation term in Eq.~\eqref{eq:Hhyb}. For $g=0$, the impurity therefore hybridises only with the even conduction band channel, while the bosonic mode activates the coupling to the odd parity channel.

{\bf Low-energy Hamiltonian --} Without the coupling to the bosonic mode, the low-energy Hamiltonian $\hat H_{\text{eff}}$ of the system in Eq.~\eqref{eq:H} for $U\gg V$ can be derived by using a conventional Schrieffer-Wolff transformation ($V$ is a $\mathbf{k}$-averaged tunnelling in the absence of the bosonic mode): Empty and doubly occupied impurity states are projected out, and we are left with the impurity spin, which is exchange-coupled to the even parity conduction band channel via the coupling constant $J={8\lvert V\rvert}^2/ \epsilon_d $ due to virtual tunnelling processes between impurity and superconductor. A static impurity spin, $\mathbf{\hat{S}}_d=S\mathbf{\hat{z}}$, therefore corresponds to a magnetic impurity which gives rise to intra-gap states, the YSR states. Inheriting the symmetry of the superconductor, these intra-gap states come in pairs of opposite energy $\pm E$ inside the gap $\Delta$, and have an electron $u\left(\mathbf{r}\right)$ and a hole component $v\left(\mathbf{r}\right)$. With the density of states at the Fermi energy $\rho\left(\epsilon_F\right)$ and the dimensionless constant $\beta=\pi \rho\left(\epsilon_{F}\right)JS/2$, we have \cite{Yu1965,Shiba1968,Rusinov1969}
\begin{align}
\frac{E}{\Delta} & = \frac{1-\beta^{2}}{1+\beta^{2}}.\label{eq:E}
\end{align}
Since the interaction is antiferromagnetic $J<0$ (due to $\epsilon_d<0$), an antiparallel alignment of the electron spin is favoured. For sufficiently strong exchange scattering, the bound state energies cross zero at the critical coupling constant $J_C=- \left[0.5\pi\rho\left(\epsilon_F\right)S\right]^{-1}$ which marks a quantum phase transition (QPT). Beyond this point $\left(J/J_C > 1\text{, }E<0\right)$, the ground state (BCS state with free impurity spin $\mathbf{S}_d=S\mathbf{\hat{z}}$) and the excited state (YSR state) interchange roles. The impurity then localises a spin-down quasiparticle in the ground state, which partially screens the impurity spin $\mathbf{S}_d=\left(S-1/2\right)\mathbf{\hat{z}}$, leaving one unpaired electron in the superconductor \cite{Sakurai1970,Balatsky2006}. A ferromagnetic interaction $J>0$, which will be encountered later, favours a parallel spin alignment, so that for $J>\lvert J_C\rvert$ the localised quasiparticle increases the impurity spin, $\mathbf{S}_d=\left(S+1/2\right)\mathbf{\hat{z}}$.

When the bosonic mode activates the coupling to the odd parity channel $\left(-\right)$, we obtain two pairs of YSR  states, one in each channel. For a high frequency $\omega_0\gg J$, we can project the low energy Hamiltonian $\hat{H}_{\text{eff}}$ onto a fixed boson number $n \ge 0$, $\hat{H}_{\text{eff}}=\hat{H}_{\text{eff}}^{n,n}\ket{n}\bra{n}$. Virtual tunnelling can now go together with the virtual absorption or emission of bosons. Since $V_{\mathbf{k},+\left(-\right)}(g\hat{Q})$ contains only even (odd) powers of $\hat{Q}$, the coupling to the even (odd) parity combination of the bath involves an even (odd) number of virtual bosons. The Hamiltonian can be split into the Hamiltonian for the even ($+$) and odd ($-$) channel, $\hat{H}^{n,n}_{\text{eff}}=\hat{H}^{n,n}_{+}+\hat{H}^{n,n}_{-}$ (see App. A),  
\begin{align} 
\hat{H}^{n,n}_{\gamma} & = \sum_{\mathbf{k}\sigma}E_{\mathbf{k}}\hat{\alpha}_{\mathbf{k}\sigma ,\gamma}^{\dagger}\hat{\alpha}_{\mathbf{k}\sigma, \gamma} -\sum_{\mathbf{k}\mathbf{k}'\gamma}J^{\mathbf{k}\mathbf{k}'}_{n,\gamma}\mathbf{\hat{S}}^{\mathbf{k}\mathbf{k}'}_{\gamma}\cdot\mathbf{\hat{S}}_d\ ,\label{eq:Heff_eo}\\
J^{\mathbf{k}\mathbf{k}'}_{n,+\left(-\right)}& = \sum_{l=\text{even(odd)}}\frac{8}{\epsilon_d -l\omega_0}V_{\mathbf{k},+\left(-\right)}^{n,n+l}\left(V_{\mathbf{k}',+\left(-\right)}^{n,n+l}\right)^{\ast}\label{eq:Jnn}.
\end{align}
Here, $\mathbf{\hat{S}}^{\mathbf{k}\mathbf{k}'}_{\gamma}=\tfrac12\sum_{\sigma\sigma'}\hat{a}_{\mathbf{k}\sigma,\gamma}^{\dagger}\bm{\sigma}_{\sigma\sigma'}\hat{a}_{\mathbf{k'}\sigma',\gamma}$ is the spin of conduction electrons with Pauli vector $\bm{\sigma}=(\sigma_x,\sigma_y,\sigma_z)$, and the exchange interaction with the coupling constant $J^{\mathbf{k}\mathbf{k}'}_{n,\gamma}$ is given in terms of the matrix elements $V_{\mathbf{k},\gamma}^{n,n+l}=\bra{n}V_{\mathbf{k},\gamma}(g\hat{Q})\ket{n+l}$; $l\geq -n$ is the number of absorbed ($l<0$) and emitted ($l>0$) bosons in the intermediate state. For $g=0$, the impurity couples only to the even channel, since the odd channel requires the exchange of at least one virtual boson.
 
{\bf Results --} 
In order to obtain quantitative results, we consider a Peierls-type coupling, for which the matrix element in Eq.~\eqref{eq:VCh} takes the form
\begin{align}
V_{\gamma}=\frac{V}{2}\left[e^{-ig\left(\hat{b}^{\dagger}+\hat{b}\right)}+\gamma e^{ig\left(\hat{b}^{\dagger}+\hat{b}\right)}\right].\label{eq:Peierls}
\end{align}
This model describes a quantum dot which is coupled to a left (L) and right (R) lead, where left and right are defined by the polarisation direction $\left(\mathbf{k}\text{,} -\mathbf{k}\rightarrow L\text{, } R\right)$. It qualitatively (up to neglecting the $\mathbf{k}$-dependence of the matrix elements) accounts for an embedded impurity in a crystal, coupled to a linearly polarised photon mode. The Peierls coupling accounts for all symmetry-allowed multi-photon processes, giving resonances at $l\omega_0=\epsilon_{d}$. A restriction to single-photon processes would eliminate resonances in the even channel and all resonances $l>1$ in the odd channel. Here, the exchange constant in Eq.~\eqref{eq:Jnn} then reads (see App. A)
\begin{align}
J_{n,+\left(-\right)} = J \sum_{l=\text{even}\left(\text{odd}\right)} \frac{j_{n,n+l}^2}{1-\frac{l\omega_{0}}{\epsilon_{d}}},\label{eq:JnnP}
\end{align}
with the matrix element $\bra{n}e^{i\gamma g\left(\hat{b}+\hat{b}^{\dagger}\right)}\ket{n+l} =i^{\lvert l \rvert}\gamma^{l}j_{n,n+l}$. The latter decays rapidly with $\left|l \right|$, such that the coupling between different photon bands decays quickly to zero \cite{Li2020}. The contribution to the exchange scattering is antiferromagnetic (ferromagnetic) if the intermediate state lies higher (lower) in energy than the ground state, corresponding to a negative (positive) value of the energy $\epsilon_d-l\omega_0$. Thus, the interaction changes its sign by traversing the resonance at $\epsilon_d=l\omega_0$. For the vacuum $\left(n=0\right)$, only the emission of virtual photons is allowed ($l>0$), so that resonances are absent (note that $\epsilon_d<0$) and the exchange interaction in both channels remains antiferromagnetic with $J_{0,\gamma}/J\leq1$. With a single photon in the cavity $\left(n=1\right)$, we can control the sign and strength of the interaction in the odd channel by traversing the resonance at $\omega_0=-\epsilon_d$. A sign change in the even channel is still not possible due to the missing two-photon resonance at $\omega_0=-\epsilon_d/2$ for $l\ge -1$.
 \begin{figure}[tbp]
\centering
\includegraphics[width=0.99\columnwidth]{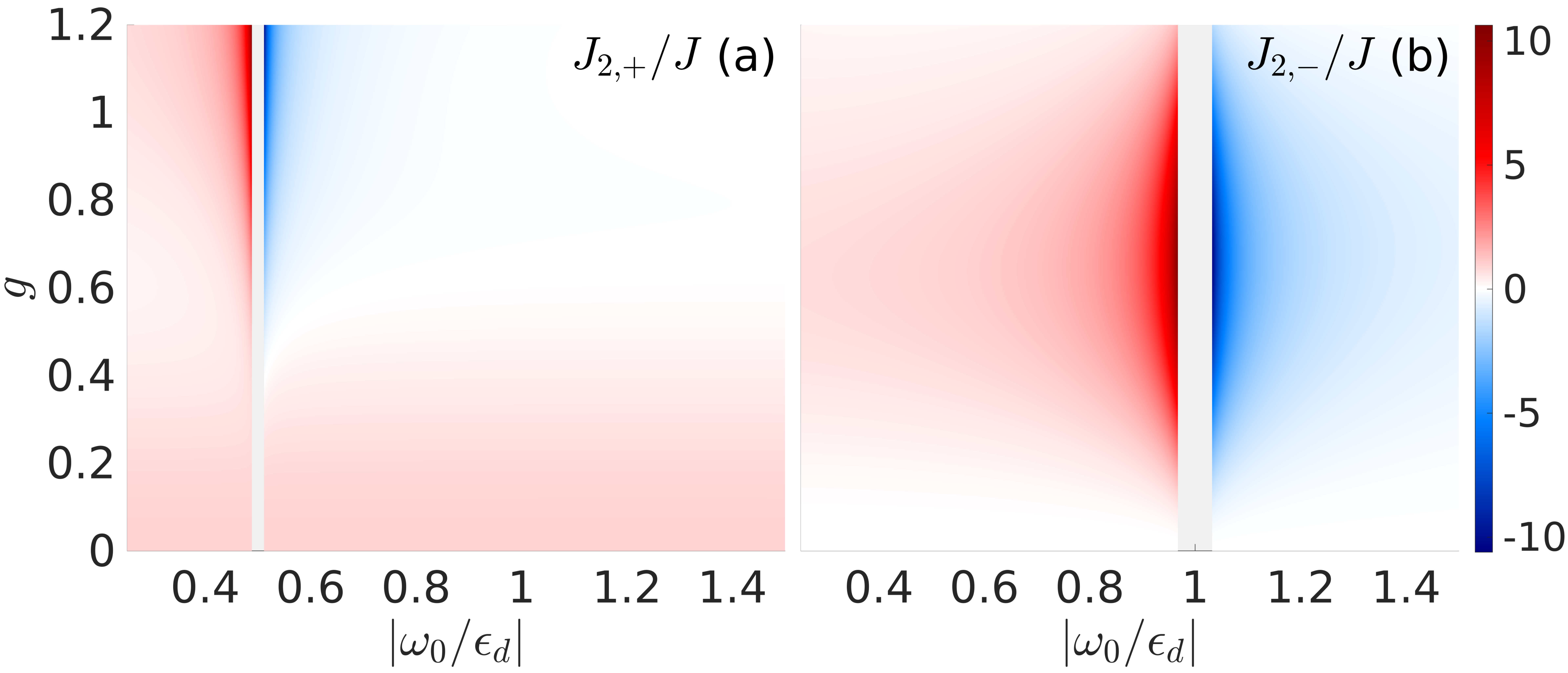}
\caption{Exchange coupling constant for the even (a) and odd (b) channel normalised by $J<0$ as a function of $g$ and $\left| \omega_{0} / \epsilon_{d} \right|$ for $n=2$. Antiferromagnetic (ferromagnetic) exchange scattering is coloured in red (blue). The cavity-uncoupled case corresponds to $J_{2,+}=1$ (red) and $J_{2,-}=0$ (white). Areas around the resonances ($\left| \omega_{0} / \epsilon_{d} \right|=1/2$, $\left| \omega_{0} / \epsilon_{d} \right|=1$) are excluded.}
	\label{fig:J}
\end{figure}
Full tunability of the exchange interaction in both channels can be already obtained by injecting two photons in the cavity. Figure~\ref{fig:J} shows the two-photon exchange constants $J_{2,\gamma}$ as a function of $\left| \omega_{0} / \epsilon_{d} \right|$ and $g$; red (blue) corresponds to antiferromagnetic (ferromagnetic) couplings. For $g=0$, the impurity couples only to the even channel, $J_{2,-}=0$, while $J_{2,+}$ reduces to the bare exchange constant $J<0$. For $g>0$, the coupling to the odd channel opens up. In the even (odd) channel, a ferromagnetic contribution is obtained right above the two-photon resonance $2\omega_0=\lvert \epsilon_d \rvert$ (one-photon resonance $\omega_0=\lvert \epsilon_d \rvert$), where one negative energy denominator $1/(\epsilon_d-l\omega_0)$ dominates. For the even channel antiferromagnetism dominates in most regions away from the resonance due to virtual tunnelling without the exchange of any photon $\left(l=0\right)$. It is enhanced below and close to the resonance, and it is weakened by the ferromagnetic contribution for larger values of $\omega_0$.
 \begin{figure}[tbp]
\centering
\includegraphics[width=0.99\columnwidth]{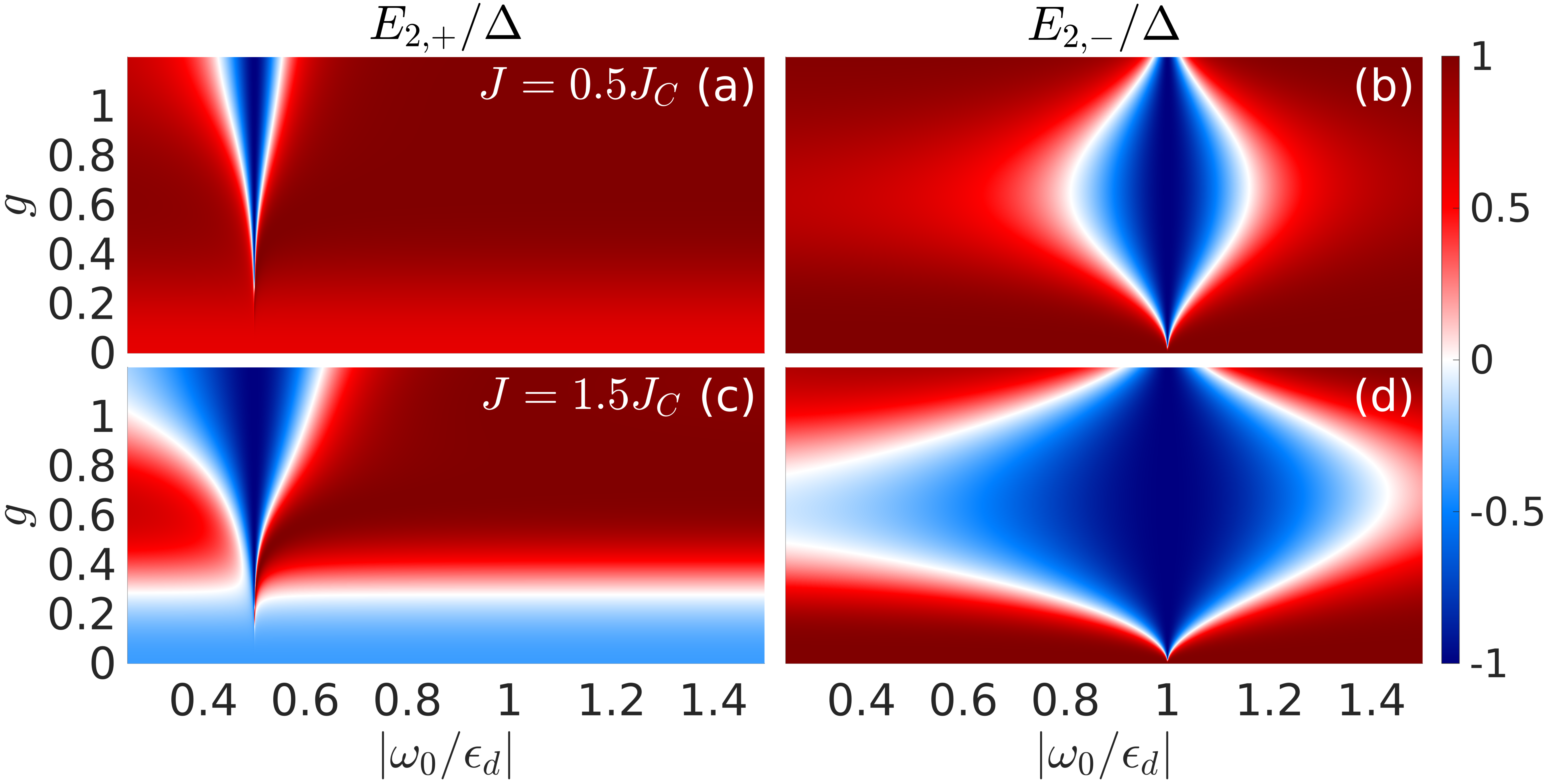}
\caption{Energy of the bound state for the even $E_{2,+}$ and odd $E_{2,-}$ channel normalised by $\Delta$ for $J=0.5 J_C $ (a)-(b) and $J=1.5 J_C $ (c)-(d) as a function of $g$ and $\left| \omega_{0} / \epsilon_{d} \right|$ for $n=2$. The Fermi energy $\mu$ is set to zero.}
	\label{E}
\end{figure}

In the context of YSR states, the strength of the interactions controls their energies, whereas the sign of the exchange coupling constant determines the spin-polarisation of the bound quasiparticles. Figure~\ref{E} shows the energies $E_{2,\gamma}$ in the even and odd channel (Eq.~\eqref{eq:E} with the couplings $J_{2,\gamma}$), for a bare exchange constant below the critical coupling $J/ J_C <1$ (taking $J_C<0)$ (see (a),(b)) and above the critical coupling ($J/ J_C >1$, (c),(d)). With the coupling to the mode $\left(g>0\right)$, in both cases we can tune the energy of the YSR states in both channels to any value within the gap $\Delta$, and also into the QPT ($E_{2,\gamma}=0$) where ground and excited state interchange roles. For $J/ J_C >1$, the reduction of exchange scattering in the even channel further away from the resonance pushes its energy towards the QPT at around $g=0.3$ and finally to $+\Delta$ (Fig.~\ref{E} (c)).

\begin{figure}[t]
\centering
\includegraphics[width=0.99\columnwidth]{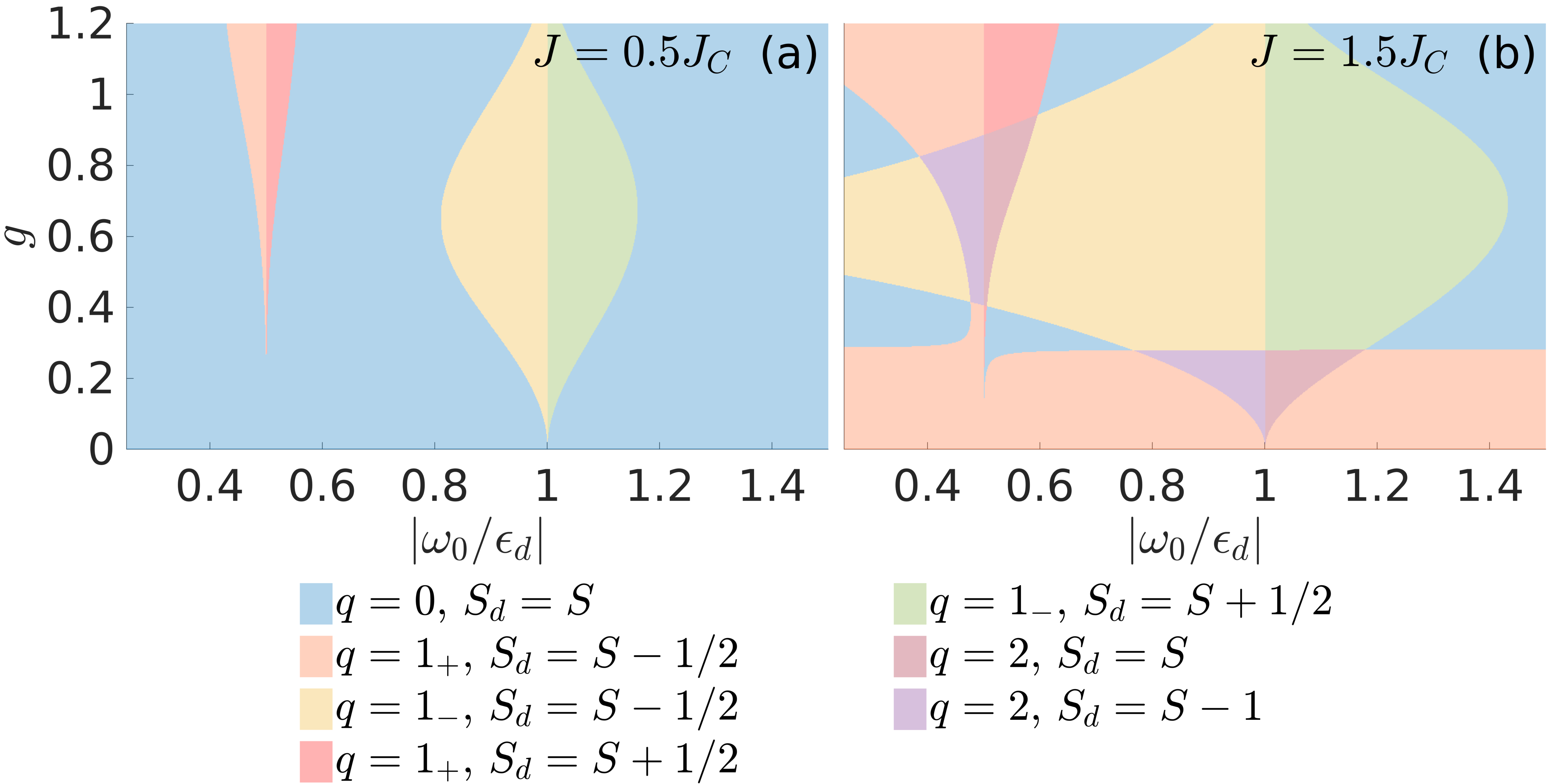}
\caption{Ground state diagram for $J/J_C=0.5$ (a) and $J/J_C=1.5$ (b) as a function of $g$ and $\left| \omega_{0} / \epsilon_{d} \right|$ for $n=2$. $q$ denotes the number of bound quasiparticle(s) within each channel and $S_d$ the impurity spin.}
	\label{GS}
\end{figure}

These observations can be easily transferred to distinguish between different ground states, which we label by the impurity spin $\mathbf{S}_d=S_d\mathbf{\hat{z}}$ and the number of bound quasiparticles within each channel $q=0,1_{+},1_{-},2$ (see Fig.~\ref{GS}), where each channel can bind up to one quasiparticle. For $J/ J_C <1$ (a), each channel can selectively undergo a QPT close to the respective resonance. Depending on the sign of $J_{2,\gamma}$, the bound quasiparticle either partially screens the impurity spin or aligns with it, which is indicated by $S_d=S-1/2$ and $S_d=S+1/2$, respectively. Above $J_C$ (b), regions open up in which both channels are simultaneously at or beyond its QPT $\left(q=2\right)$. For antiferromagnetic exchange scattering in both channels, the impurity spin is reduced to $S_d=S-1$, whereas for opposite interaction types the impurity spin appears to be free ($S_d=S$). The ground state in which two quasiparticles align parallel to the impurity ($S_d=S+1$) is absent, due to the reduced ferromagnetic exchange scattering in the even channel.

A significant effect of the mode on the exchange interactions requires a sufficiently strong coupling $g$. For the Peierls coupling, $g$ is the dipole energy normalised by the photon energy. Hence, for a free space mode wavelength $\lambda$ of $10^{-6}$ m and a tunnelling distance $a$ of $0.5$ nm, reaching the coupling $g=0.1$ requires the normalised mode volume $V_{\text{mode}}/\lambda^3=1.82\cdot 10^{-7}$ (using $\sqrt{\left(\hbar \omega_0\right)/\left(2\epsilon_0 V_{\text{mode}}\right)}$ for the vacuum electric field). Such small mode volumes may be achieved in nanoplasmonic cavities, where the light field is highly confined and strongly enhanced at nanometer-sized structures. For the cavity control of quantum impurity physics, a particularly interesting direction would be to employ plasmonic picocavities based on AFM/STM tips \cite{Maccaferri2021}, which provide a tunable cavity setting. Such picocavities have already been used to strongly couple atomically confined light to vibrational modes of single molecules (see, e.g., \cite{Benz2016,Lee2019}), and to exciton-based single quantum emitters \cite{Park2019}, with a volume compression $V_{\text{mode}}/\lambda^3$ below $10^{-6}$. Moving towards lower photon frequencies allows larger mode volumes but results in a reduced controllability within the present setting, since the largest effect on the exchange couplings is obtained for near resonant couplings.

An alternative route to control the couplings is to consider a classically driven system, where the strength of the light-matter interaction can be controlled by the amplitude of the drive. In this classical limit, the state of the light field is unaffected by the solid and $g\hat{Q}$ is replaced by a time-dependent field $A\cos\left(\omega_{0}t\right)$, with dimensionless amplitude $A$. This gives rise to a time-periodic Hamiltonian $\hat{H}\left(t\right)=\hat{H}\left(t+T\right)$ with period $T=2\pi/\omega_0$. The stroboscopic dynamics, i.e., the evolution of the system over one period of the drive, is described by a time-independent Hamiltonian, the Floquet Hamiltonian $\hat{H}_F$. The eigenstates, i.e., Floquet states, with their quasienergies $\epsilon$ as eigenvalues can be interpreted as quantum states of a definite but very large number of photons \cite{Shirley1965,Sentef2020,Li2020}. Thus, it is not surprising that the Floquet Hamiltonian emerges from the quantum formalism in the limit $g\rightarrow 0$ and a photon number $n\to \infty$, such that $2g\sqrt{n}=A$ is a fixed number. This correspondence flows into a similarity between the results for the classically driven system, shown in App. B, and the two-photon system, since in addition the coupling between different photon bands decays quickly to zero. An exception is apparent due to the infinite number of photons in the classical field, since in this case all resonances (at $l\omega_{0}=\epsilon_d$) are present. The energy of the YSR states is now the quasienergy with the corresponding Floquet state. Such a Floquet state can be reached adiabatically by starting from the stationary bound state in equilibrium and turning on the drive sufficiently slowly, such that the state evolves within the same quasienergy. Heating, which has been identified as a potential experimental obstacle in bulk Floquet systems, might be mitigated for the impurity setting investigated in this work, since the host material will act as a bath \cite{Eckstein2017}.

In conclusion, for an Anderson impurity embedded in a conventional superconductor, we showed that a bosonic mode activates the coupling to the p-wave conduction band channel in addition to the s-wave channel. For inversion symmetry around the impurity, these two channels are independent, and two pairs of YSR states emerge. The exchange couplings in the two channels depend sensitively on the state of the bosonic mode, such that the nature of the YSR states can be controlled widely by varying the amplitude and frequency in the limit of weak coupling and strong classical drive (Floquet engineering), or by the precise quantum state of the bosonic mode in the strongly coupled limit. For strong coupling, already the presence of only two bosons gives a degree of controllability which is comparable to classical Floquet engineering. The strongly coupled quantum case could be realised by placing the YSR impurity inside a nanoplasmonic cavity. The bosonic mode is either the cavity field itself, or alternatively the cavity can hybridise with a vibrational mode, which in turn couples to the electronic transitions between impurity and superconductor. While we have computed the exchange interactions in the presence of bosonic Fock states, in an experiment the cavity would be excited into the multi-photon regime by driving with an external laser field. The strong nonlinearity which we have demonstrated for the Fock states, with a potentially different ground state of the impurity depending on the number of bosons, indicates that exploring various driving protocols and the resulting intertwined dynamics of cavity photons and the YSR states will be an interesting direction for future research, including the generation of non-classical light. Moreover, the interplay between gain and dissipation in a cavity leading to a non-hermitian system is worth to be investigated. Beyond that, we expect our results to be relevant for clusters of YSR impurities giving rise to unconventional superconductivity and Majorana bound states, where the combination of collective effects with the possibility of controlling the coupling between YSR states could lead to interesting topological phenomena. Finally, the modification of the light field could be used as a sensitive probe of the solid state system. 

\begin{acknowledgments}
M. E. was  funded by the ERC Starting Grant No. 716648. S.V.K. acknowledge funding by the Max Planck Society in the form of a Max Planck Research Group. M.E. and S.V.K. acknowledge funding by the Deutsche Forschungsgemeinschaft (DFG, German Research Foundation) -- Project-ID 429529648 -- TRR 306 QuCoLiMa (``Quantum Cooperativity of Light and Matter'').

\end{acknowledgments}

\bibliography{mybib}

\begin{thebibliography}{58}%
\makeatletter
\providecommand \@ifxundefined [1]{%
 \@ifx{#1\undefined}
}%
\providecommand \@ifnum [1]{%
 \ifnum #1\expandafter \@firstoftwo
 \else \expandafter \@secondoftwo
 \fi
}%
\providecommand \@ifx [1]{%
 \ifx #1\expandafter \@firstoftwo
 \else \expandafter \@secondoftwo
 \fi
}%
\providecommand \natexlab [1]{#1}%
\providecommand \enquote  [1]{``#1''}%
\providecommand \bibnamefont  [1]{#1}%
\providecommand \bibfnamefont [1]{#1}%
\providecommand \citenamefont [1]{#1}%
\providecommand \href@noop [0]{\@secondoftwo}%
\providecommand \href [0]{\begingroup \@sanitize@url \@href}%
\providecommand \@href[1]{\@@startlink{#1}\@@href}%
\providecommand \@@href[1]{\endgroup#1\@@endlink}%
\providecommand \@sanitize@url [0]{\catcode `\\12\catcode `\$12\catcode
  `\&12\catcode `\#12\catcode `\^12\catcode `\_12\catcode `\%12\relax}%
\providecommand \@@startlink[1]{}%
\providecommand \@@endlink[0]{}%
\providecommand \url  [0]{\begingroup\@sanitize@url \@url }%
\providecommand \@url [1]{\endgroup\@href {#1}{\urlprefix }}%
\providecommand \urlprefix  [0]{URL }%
\providecommand \Eprint [0]{\href }%
\providecommand \doibase [0]{https://doi.org/}%
\providecommand \selectlanguage [0]{\@gobble}%
\providecommand \bibinfo  [0]{\@secondoftwo}%
\providecommand \bibfield  [0]{\@secondoftwo}%
\providecommand \translation [1]{[#1]}%
\providecommand \BibitemOpen [0]{}%
\providecommand \bibitemStop [0]{}%
\providecommand \bibitemNoStop [0]{.\EOS\space}%
\providecommand \EOS [0]{\spacefactor3000\relax}%
\providecommand \BibitemShut  [1]{\csname bibitem#1\endcsname}%
\let\auto@bib@innerbib\@empty
\bibitem [{\citenamefont {Yu}(1965)}]{Yu1965}%
  \BibitemOpen
  \bibfield  {author} {\bibinfo {author} {\bibfnamefont {L.}~\bibnamefont
  {Yu}},\ }\href {http://en.cnki.com.cn/Article_en/CJFDTOTAL-WLXB196501007.htm}
  {\bibfield  {journal} {\bibinfo  {journal} {Acta Phys. Sin.}\ }\textbf
  {\bibinfo {volume} {21}} (\bibinfo {year} {1965})}\BibitemShut {NoStop}%
\bibitem [{\citenamefont {Shiba}(1968)}]{Shiba1968}%
  \BibitemOpen
  \bibfield  {author} {\bibinfo {author} {\bibfnamefont {H.}~\bibnamefont
  {Shiba}},\ }\href {https://doi.org/10.1143/PTP.40.435} {\bibfield  {journal}
  {\bibinfo  {journal} {Prog. Theor. Phys.}\ }\textbf {\bibinfo {volume}
  {40}},\ \bibinfo {pages} {435} (\bibinfo {year} {1968})}\BibitemShut
  {NoStop}%
\bibitem [{\citenamefont {Rusinov}(1969)}]{Rusinov1969}%
  \BibitemOpen
  \bibfield  {author} {\bibinfo {author} {\bibfnamefont {A.~I.}\ \bibnamefont
  {Rusinov}},\ }\href {http://jetpletters.ru/ps/1658/article_25295.shtml}
  {\bibfield  {journal} {\bibinfo  {journal} {JETP Lett.}\ }\textbf {\bibinfo
  {volume} {9}},\ \bibinfo {pages} {85} (\bibinfo {year} {1969})}\BibitemShut
  {NoStop}%
\bibitem [{\citenamefont {Yazdani}\ \emph {et~al.}(1997)\citenamefont
  {Yazdani}, \citenamefont {Jones}, \citenamefont {Lutz}, \citenamefont
  {Crommie},\ and\ \citenamefont {Eigler}}]{Yazdani1997}%
  \BibitemOpen
  \bibfield  {author} {\bibinfo {author} {\bibfnamefont {A.}~\bibnamefont
  {Yazdani}}, \bibinfo {author} {\bibfnamefont {B.~A.}\ \bibnamefont {Jones}},
  \bibinfo {author} {\bibfnamefont {C.~P.}\ \bibnamefont {Lutz}}, \bibinfo
  {author} {\bibfnamefont {M.~F.}\ \bibnamefont {Crommie}},\ and\ \bibinfo
  {author} {\bibfnamefont {D.~M.}\ \bibnamefont {Eigler}},\ }\href
  {https://doi.org/10.1126/science.275.5307.1767} {\bibfield  {journal}
  {\bibinfo  {journal} {Science}\ }\textbf {\bibinfo {volume} {275}},\ \bibinfo
  {pages} {1767} (\bibinfo {year} {1997})}\BibitemShut {NoStop}%
\bibitem [{\citenamefont {Balatsky}\ \emph {et~al.}(2006)\citenamefont
  {Balatsky}, \citenamefont {Vekhter},\ and\ \citenamefont
  {Zhu}}]{Balatsky2006}%
  \BibitemOpen
  \bibfield  {author} {\bibinfo {author} {\bibfnamefont {A.~V.}\ \bibnamefont
  {Balatsky}}, \bibinfo {author} {\bibfnamefont {I.}~\bibnamefont {Vekhter}},\
  and\ \bibinfo {author} {\bibfnamefont {J.-X.}\ \bibnamefont {Zhu}},\ }\href
  {https://doi.org/10.1103/RevModPhys.78.373} {\bibfield  {journal} {\bibinfo
  {journal} {Rev. Mod. Phys.}\ }\textbf {\bibinfo {volume} {78}},\ \bibinfo
  {pages} {373} (\bibinfo {year} {2006})}\BibitemShut {NoStop}%
\bibitem [{\citenamefont {Choy}\ \emph {et~al.}(2011)\citenamefont {Choy},
  \citenamefont {Edge}, \citenamefont {Akhmerov},\ and\ \citenamefont
  {Beenakker}}]{Choy2011}%
  \BibitemOpen
  \bibfield  {author} {\bibinfo {author} {\bibfnamefont {T.-P.}\ \bibnamefont
  {Choy}}, \bibinfo {author} {\bibfnamefont {J.~M.}\ \bibnamefont {Edge}},
  \bibinfo {author} {\bibfnamefont {A.~R.}\ \bibnamefont {Akhmerov}},\ and\
  \bibinfo {author} {\bibfnamefont {C.~W.~J.}\ \bibnamefont {Beenakker}},\
  }\href {https://doi.org/10.1103/PhysRevB.84.195442} {\bibfield  {journal}
  {\bibinfo  {journal} {Phys. Rev. B}\ }\textbf {\bibinfo {volume} {84}},\
  \bibinfo {pages} {195442} (\bibinfo {year} {2011})}\BibitemShut {NoStop}%
\bibitem [{\citenamefont {Nakosai}\ \emph {et~al.}(2013)\citenamefont
  {Nakosai}, \citenamefont {Tanaka},\ and\ \citenamefont
  {Nagaosa}}]{Nakosai2013}%
  \BibitemOpen
  \bibfield  {author} {\bibinfo {author} {\bibfnamefont {S.}~\bibnamefont
  {Nakosai}}, \bibinfo {author} {\bibfnamefont {Y.}~\bibnamefont {Tanaka}},\
  and\ \bibinfo {author} {\bibfnamefont {N.}~\bibnamefont {Nagaosa}},\ }\href
  {https://doi.org/10.1103/PhysRevB.88.180503} {\bibfield  {journal} {\bibinfo
  {journal} {Phys. Rev. B}\ }\textbf {\bibinfo {volume} {88}},\ \bibinfo
  {pages} {180503} (\bibinfo {year} {2013})}\BibitemShut {NoStop}%
\bibitem [{\citenamefont {Nadj-Perge}\ \emph {et~al.}(2013)\citenamefont
  {Nadj-Perge}, \citenamefont {Drozdov}, \citenamefont {Bernevig},\ and\
  \citenamefont {Yazdani}}]{Perge2013}%
  \BibitemOpen
  \bibfield  {author} {\bibinfo {author} {\bibfnamefont {S.}~\bibnamefont
  {Nadj-Perge}}, \bibinfo {author} {\bibfnamefont {I.~K.}\ \bibnamefont
  {Drozdov}}, \bibinfo {author} {\bibfnamefont {B.~A.}\ \bibnamefont
  {Bernevig}},\ and\ \bibinfo {author} {\bibfnamefont {A.}~\bibnamefont
  {Yazdani}},\ }\href {https://doi.org/10.1103/PhysRevB.88.020407} {\bibfield
  {journal} {\bibinfo  {journal} {Phys. Rev. B}\ }\textbf {\bibinfo {volume}
  {88}},\ \bibinfo {pages} {020407} (\bibinfo {year} {2013})}\BibitemShut
  {NoStop}%
\bibitem [{\citenamefont {Nadj-Perge}\ \emph {et~al.}(2014)\citenamefont
  {Nadj-Perge}, \citenamefont {Drozdov}, \citenamefont {Li}, \citenamefont
  {Chen}, \citenamefont {Jeon}, \citenamefont {Seo}, \citenamefont {MacDonald},
  \citenamefont {Bernevig},\ and\ \citenamefont {Yazdani}}]{Nadj-Perge2014}%
  \BibitemOpen
  \bibfield  {author} {\bibinfo {author} {\bibfnamefont {S.}~\bibnamefont
  {Nadj-Perge}}, \bibinfo {author} {\bibfnamefont {I.~K.}\ \bibnamefont
  {Drozdov}}, \bibinfo {author} {\bibfnamefont {J.}~\bibnamefont {Li}},
  \bibinfo {author} {\bibfnamefont {H.}~\bibnamefont {Chen}}, \bibinfo {author}
  {\bibfnamefont {S.}~\bibnamefont {Jeon}}, \bibinfo {author} {\bibfnamefont
  {J.}~\bibnamefont {Seo}}, \bibinfo {author} {\bibfnamefont {A.~H.}\
  \bibnamefont {MacDonald}}, \bibinfo {author} {\bibfnamefont {B.~A.}\
  \bibnamefont {Bernevig}},\ and\ \bibinfo {author} {\bibfnamefont
  {A.}~\bibnamefont {Yazdani}},\ }\href
  {https://doi.org/10.1126/science.1259327} {\bibfield  {journal} {\bibinfo
  {journal} {Science}\ }\textbf {\bibinfo {volume} {346}},\ \bibinfo {pages}
  {602} (\bibinfo {year} {2014})}\BibitemShut {NoStop}%
\bibitem [{\citenamefont {Perrin}\ \emph {et~al.}(2020)\citenamefont {Perrin},
  \citenamefont {Santos}, \citenamefont {M\'enard}, \citenamefont {Brun},
  \citenamefont {Cren}, \citenamefont {Civelli},\ and\ \citenamefont
  {Simon}}]{Perrin2020}%
  \BibitemOpen
  \bibfield  {author} {\bibinfo {author} {\bibfnamefont {V.}~\bibnamefont
  {Perrin}}, \bibinfo {author} {\bibfnamefont {F.~L.~N.}\ \bibnamefont
  {Santos}}, \bibinfo {author} {\bibfnamefont {G.~C.}\ \bibnamefont
  {M\'enard}}, \bibinfo {author} {\bibfnamefont {C.}~\bibnamefont {Brun}},
  \bibinfo {author} {\bibfnamefont {T.}~\bibnamefont {Cren}}, \bibinfo {author}
  {\bibfnamefont {M.}~\bibnamefont {Civelli}},\ and\ \bibinfo {author}
  {\bibfnamefont {P.}~\bibnamefont {Simon}},\ }\href
  {https://doi.org/10.1103/PhysRevLett.125.117003} {\bibfield  {journal}
  {\bibinfo  {journal} {Phys. Rev. Lett.}\ }\textbf {\bibinfo {volume} {125}},\
  \bibinfo {pages} {117003} (\bibinfo {year} {2020})}\BibitemShut {NoStop}%
\bibitem [{\citenamefont {Schneider}\ \emph {et~al.}(2021)\citenamefont
  {Schneider}, \citenamefont {Beck}, \citenamefont {Posske}, \citenamefont
  {Crawford}, \citenamefont {Mascot}, \citenamefont {Rachel}, \citenamefont
  {Wiesendanger},\ and\ \citenamefont {Wiebe}}]{Schneider2021}%
  \BibitemOpen
  \bibfield  {author} {\bibinfo {author} {\bibfnamefont {L.}~\bibnamefont
  {Schneider}}, \bibinfo {author} {\bibfnamefont {P.}~\bibnamefont {Beck}},
  \bibinfo {author} {\bibfnamefont {T.}~\bibnamefont {Posske}}, \bibinfo
  {author} {\bibfnamefont {D.}~\bibnamefont {Crawford}}, \bibinfo {author}
  {\bibfnamefont {E.}~\bibnamefont {Mascot}}, \bibinfo {author} {\bibfnamefont
  {S.}~\bibnamefont {Rachel}}, \bibinfo {author} {\bibfnamefont
  {R.}~\bibnamefont {Wiesendanger}},\ and\ \bibinfo {author} {\bibfnamefont
  {J.}~\bibnamefont {Wiebe}},\ }\href
  {https://doi.org/10.1038/s41567-021-01234-y} {\bibfield  {journal} {\bibinfo
  {journal} {Nat. Phys.}\ }\textbf {\bibinfo {volume} {17}},\ \bibinfo {pages}
  {943} (\bibinfo {year} {2021})}\BibitemShut {NoStop}%
\bibitem [{\citenamefont {Akkaravarawong}\ \emph {et~al.}(2019)\citenamefont
  {Akkaravarawong}, \citenamefont {V\"ayrynen}, \citenamefont {Sau},
  \citenamefont {Demler}, \citenamefont {Glazman},\ and\ \citenamefont
  {Yao}}]{Akkaravarawong2019}%
  \BibitemOpen
  \bibfield  {author} {\bibinfo {author} {\bibfnamefont {K.}~\bibnamefont
  {Akkaravarawong}}, \bibinfo {author} {\bibfnamefont {J.~I.}\ \bibnamefont
  {V\"ayrynen}}, \bibinfo {author} {\bibfnamefont {J.~D.}\ \bibnamefont {Sau}},
  \bibinfo {author} {\bibfnamefont {E.~A.}\ \bibnamefont {Demler}}, \bibinfo
  {author} {\bibfnamefont {L.~I.}\ \bibnamefont {Glazman}},\ and\ \bibinfo
  {author} {\bibfnamefont {N.~Y.}\ \bibnamefont {Yao}},\ }\href
  {https://doi.org/10.1103/PhysRevResearch.1.033091} {\bibfield  {journal}
  {\bibinfo  {journal} {Phys. Rev. Research}\ }\textbf {\bibinfo {volume}
  {1}},\ \bibinfo {pages} {033091} (\bibinfo {year} {2019})}\BibitemShut
  {NoStop}%
\bibitem [{\citenamefont {Franke}\ \emph {et~al.}(2011)\citenamefont {Franke},
  \citenamefont {Schulze},\ and\ \citenamefont {Pascual}}]{Franke2011}%
  \BibitemOpen
  \bibfield  {author} {\bibinfo {author} {\bibfnamefont {K.~J.}\ \bibnamefont
  {Franke}}, \bibinfo {author} {\bibfnamefont {G.}~\bibnamefont {Schulze}},\
  and\ \bibinfo {author} {\bibfnamefont {J.~I.}\ \bibnamefont {Pascual}},\
  }\href {https://doi.org/10.1126/science.1202204} {\bibfield  {journal}
  {\bibinfo  {journal} {Science}\ }\textbf {\bibinfo {volume} {332}},\ \bibinfo
  {pages} {940} (\bibinfo {year} {2011})}\BibitemShut {NoStop}%
\bibitem [{\citenamefont {Farinacci}\ \emph {et~al.}(2018)\citenamefont
  {Farinacci}, \citenamefont {Ahmadi}, \citenamefont {Reecht}, \citenamefont
  {Ruby}, \citenamefont {Bogdanoff}, \citenamefont {Peters}, \citenamefont
  {Heinrich}, \citenamefont {von Oppen},\ and\ \citenamefont
  {Franke}}]{Farinacci2018}%
  \BibitemOpen
  \bibfield  {author} {\bibinfo {author} {\bibfnamefont {L.}~\bibnamefont
  {Farinacci}}, \bibinfo {author} {\bibfnamefont {G.}~\bibnamefont {Ahmadi}},
  \bibinfo {author} {\bibfnamefont {G.}~\bibnamefont {Reecht}}, \bibinfo
  {author} {\bibfnamefont {M.}~\bibnamefont {Ruby}}, \bibinfo {author}
  {\bibfnamefont {N.}~\bibnamefont {Bogdanoff}}, \bibinfo {author}
  {\bibfnamefont {O.}~\bibnamefont {Peters}}, \bibinfo {author} {\bibfnamefont
  {B.~W.}\ \bibnamefont {Heinrich}}, \bibinfo {author} {\bibfnamefont
  {F.}~\bibnamefont {von Oppen}},\ and\ \bibinfo {author} {\bibfnamefont
  {K.~J.}\ \bibnamefont {Franke}},\ }\href
  {https://doi.org/10.1103/PhysRevLett.121.196803} {\bibfield  {journal}
  {\bibinfo  {journal} {Phys. Rev. Lett.}\ }\textbf {\bibinfo {volume} {121}},\
  \bibinfo {pages} {196803} (\bibinfo {year} {2018})}\BibitemShut {NoStop}%
\bibitem [{\citenamefont {Nakagawa}\ and\ \citenamefont
  {Kawakami}(2015)}]{Nakagawa2015}%
  \BibitemOpen
  \bibfield  {author} {\bibinfo {author} {\bibfnamefont {M.}~\bibnamefont
  {Nakagawa}}\ and\ \bibinfo {author} {\bibfnamefont {N.}~\bibnamefont
  {Kawakami}},\ }\href {https://doi.org/10.1103/PhysRevLett.115.165303}
  {\bibfield  {journal} {\bibinfo  {journal} {Phys. Rev. Lett.}\ }\textbf
  {\bibinfo {volume} {115}},\ \bibinfo {pages} {165303} (\bibinfo {year}
  {2015})}\BibitemShut {NoStop}%
\bibitem [{\citenamefont {Iwahori}\ and\ \citenamefont
  {Kawakami}(2016)}]{Iwahori2016}%
  \BibitemOpen
  \bibfield  {author} {\bibinfo {author} {\bibfnamefont {K.}~\bibnamefont
  {Iwahori}}\ and\ \bibinfo {author} {\bibfnamefont {N.}~\bibnamefont
  {Kawakami}},\ }\href {https://doi.org/10.1103/PhysRevA.94.063647} {\bibfield
  {journal} {\bibinfo  {journal} {Phys. Rev. A}\ }\textbf {\bibinfo {volume}
  {94}},\ \bibinfo {pages} {063647} (\bibinfo {year} {2016})}\BibitemShut
  {NoStop}%
\bibitem [{\citenamefont {Eckstein}\ and\ \citenamefont
  {Werner}(2017)}]{Eckstein2017}%
  \BibitemOpen
  \bibfield  {author} {\bibinfo {author} {\bibfnamefont {M.}~\bibnamefont
  {Eckstein}}\ and\ \bibinfo {author} {\bibfnamefont {P.}~\bibnamefont
  {Werner}},\ }\href@noop {} {} (\bibinfo {year} {2017}),\ \Eprint
  {https://arxiv.org/abs/1704.02300} {arXiv:1704.02300 [cond-mat.str-el]}
  \BibitemShut {NoStop}%
\bibitem [{\citenamefont {Quito}\ and\ \citenamefont
  {Flint}(2021)}]{Quito2021}%
  \BibitemOpen
  \bibfield  {author} {\bibinfo {author} {\bibfnamefont {V.~L.}\ \bibnamefont
  {Quito}}\ and\ \bibinfo {author} {\bibfnamefont {R.}~\bibnamefont {Flint}},\
  }\href@noop {} {} (\bibinfo {year} {2021}),\ \Eprint
  {https://arxiv.org/abs/2111.07994} {arXiv:2111.07994 [cond-mat.str-el]}
  \BibitemShut {NoStop}%
\bibitem [{\citenamefont {Basov}\ \emph {et~al.}(2017)\citenamefont {Basov},
  \citenamefont {Averitt},\ and\ \citenamefont {Hsieh}}]{Basov2017}%
  \BibitemOpen
  \bibfield  {author} {\bibinfo {author} {\bibfnamefont {D.~N.}\ \bibnamefont
  {Basov}}, \bibinfo {author} {\bibfnamefont {R.~D.}\ \bibnamefont {Averitt}},\
  and\ \bibinfo {author} {\bibfnamefont {D.}~\bibnamefont {Hsieh}},\ }\href
  {https://doi.org/10.1038/nmat5017} {\bibfield  {journal} {\bibinfo  {journal}
  {Nat. Mater.}\ }\textbf {\bibinfo {volume} {16}},\ \bibinfo {pages} {1077}
  (\bibinfo {year} {2017})}\BibitemShut {NoStop}%
\bibitem [{\citenamefont {de~la Torre}\ \emph {et~al.}(2021)\citenamefont
  {de~la Torre}, \citenamefont {Kennes}, \citenamefont {Claassen},
  \citenamefont {Gerber}, \citenamefont {McIver},\ and\ \citenamefont
  {Sentef}}]{Torre2021}%
  \BibitemOpen
  \bibfield  {author} {\bibinfo {author} {\bibfnamefont {A.}~\bibnamefont
  {de~la Torre}}, \bibinfo {author} {\bibfnamefont {D.~M.}\ \bibnamefont
  {Kennes}}, \bibinfo {author} {\bibfnamefont {M.}~\bibnamefont {Claassen}},
  \bibinfo {author} {\bibfnamefont {S.}~\bibnamefont {Gerber}}, \bibinfo
  {author} {\bibfnamefont {J.~W.}\ \bibnamefont {McIver}},\ and\ \bibinfo
  {author} {\bibfnamefont {M.~A.}\ \bibnamefont {Sentef}},\ }\href
  {https://doi.org/10.1103/RevModPhys.93.041002} {\bibfield  {journal}
  {\bibinfo  {journal} {Rev. Mod. Phys.}\ }\textbf {\bibinfo {volume} {93}},\
  \bibinfo {pages} {041002} (\bibinfo {year} {2021})}\BibitemShut {NoStop}%
\bibitem [{\citenamefont {McIver}\ \emph {et~al.}(2020)\citenamefont {McIver},
  \citenamefont {Schulte}, \citenamefont {Stein}, \citenamefont {Matsuyama},
  \citenamefont {Jotzu}, \citenamefont {Meier},\ and\ \citenamefont
  {Cavalleri}}]{McIver2020}%
  \BibitemOpen
  \bibfield  {author} {\bibinfo {author} {\bibfnamefont {J.~W.}\ \bibnamefont
  {McIver}}, \bibinfo {author} {\bibfnamefont {B.}~\bibnamefont {Schulte}},
  \bibinfo {author} {\bibfnamefont {F.-U.}\ \bibnamefont {Stein}}, \bibinfo
  {author} {\bibfnamefont {T.}~\bibnamefont {Matsuyama}}, \bibinfo {author}
  {\bibfnamefont {G.}~\bibnamefont {Jotzu}}, \bibinfo {author} {\bibfnamefont
  {G.}~\bibnamefont {Meier}},\ and\ \bibinfo {author} {\bibfnamefont
  {A.}~\bibnamefont {Cavalleri}},\ }\href
  {https://doi.org/10.1038/s41567-019-0698-y} {\bibfield  {journal} {\bibinfo
  {journal} {Nat. Phys.}\ }\textbf {\bibinfo {volume} {16}},\ \bibinfo {pages}
  {38} (\bibinfo {year} {2020})}\BibitemShut {NoStop}%
\bibitem [{\citenamefont {Wang}\ \emph {et~al.}(2013)\citenamefont {Wang},
  \citenamefont {Steinberg}, \citenamefont {Jarillo-Herrero},\ and\
  \citenamefont {Gedik}}]{Wang2013}%
  \BibitemOpen
  \bibfield  {author} {\bibinfo {author} {\bibfnamefont {Y.~H.}\ \bibnamefont
  {Wang}}, \bibinfo {author} {\bibfnamefont {H.}~\bibnamefont {Steinberg}},
  \bibinfo {author} {\bibfnamefont {P.}~\bibnamefont {Jarillo-Herrero}},\ and\
  \bibinfo {author} {\bibfnamefont {N.}~\bibnamefont {Gedik}},\ }\href
  {https://doi.org/10.1126/science.1239834} {\bibfield  {journal} {\bibinfo
  {journal} {Science}\ }\textbf {\bibinfo {volume} {342}},\ \bibinfo {pages}
  {453} (\bibinfo {year} {2013})}\BibitemShut {NoStop}%
\bibitem [{\citenamefont {Oka}\ and\ \citenamefont {Kitamura}(2019)}]{Oka2019}%
  \BibitemOpen
  \bibfield  {author} {\bibinfo {author} {\bibfnamefont {T.}~\bibnamefont
  {Oka}}\ and\ \bibinfo {author} {\bibfnamefont {S.}~\bibnamefont {Kitamura}},\
  }\href {https://doi.org/10.1146/annurev-conmatphys-031218-013423} {\bibfield
  {journal} {\bibinfo  {journal} {Annu. Rev. Condens. Matter Phys.}\ }\textbf
  {\bibinfo {volume} {10}},\ \bibinfo {pages} {387} (\bibinfo {year}
  {2019})}\BibitemShut {NoStop}%
\bibitem [{\citenamefont {Mentink}\ \emph {et~al.}(2015)\citenamefont
  {Mentink}, \citenamefont {Balzer},\ and\ \citenamefont
  {Eckstein}}]{Mentink2015}%
  \BibitemOpen
  \bibfield  {author} {\bibinfo {author} {\bibfnamefont {J.~H.}\ \bibnamefont
  {Mentink}}, \bibinfo {author} {\bibfnamefont {K.}~\bibnamefont {Balzer}},\
  and\ \bibinfo {author} {\bibfnamefont {M.}~\bibnamefont {Eckstein}},\ }\href
  {https://doi.org/10.1038/ncomms7708} {\bibfield  {journal} {\bibinfo
  {journal} {Nat. Commun.}\ }\textbf {\bibinfo {volume} {6}},\ \bibinfo {pages}
  {6708} (\bibinfo {year} {2015})}\BibitemShut {NoStop}%
\bibitem [{\citenamefont {Bukov}\ \emph {et~al.}(2015)\citenamefont {Bukov},
  \citenamefont {D'Alessio},\ and\ \citenamefont {Polkovnikov}}]{Bukov2015}%
  \BibitemOpen
  \bibfield  {author} {\bibinfo {author} {\bibfnamefont {M.}~\bibnamefont
  {Bukov}}, \bibinfo {author} {\bibfnamefont {L.}~\bibnamefont {D'Alessio}},\
  and\ \bibinfo {author} {\bibfnamefont {A.}~\bibnamefont {Polkovnikov}},\
  }\href {https://doi.org/10.1080/00018732.2015.1055918} {\bibfield  {journal}
  {\bibinfo  {journal} {Adv. Phys.}\ }\textbf {\bibinfo {volume} {64}},\
  \bibinfo {pages} {139} (\bibinfo {year} {2015})}\BibitemShut {NoStop}%
\bibitem [{\citenamefont {Claassen}\ \emph {et~al.}(2017)\citenamefont
  {Claassen}, \citenamefont {Jiang}, \citenamefont {Moritz},\ and\
  \citenamefont {Devereaux}}]{Claassen2017}%
  \BibitemOpen
  \bibfield  {author} {\bibinfo {author} {\bibfnamefont {M.}~\bibnamefont
  {Claassen}}, \bibinfo {author} {\bibfnamefont {H.-C.}\ \bibnamefont {Jiang}},
  \bibinfo {author} {\bibfnamefont {B.}~\bibnamefont {Moritz}},\ and\ \bibinfo
  {author} {\bibfnamefont {T.~P.}\ \bibnamefont {Devereaux}},\ }\href
  {https://doi.org/https://doi.org/10.1038/s41467-017-00876-y} {\bibfield
  {journal} {\bibinfo  {journal} {Nat. Commun.}\ }\textbf {\bibinfo {volume}
  {8}},\ \bibinfo {pages} {1192} (\bibinfo {year} {2017})}\BibitemShut
  {NoStop}%
\bibitem [{\citenamefont {Sentef}\ \emph {et~al.}(2017)\citenamefont {Sentef},
  \citenamefont {Tokuno}, \citenamefont {Georges},\ and\ \citenamefont
  {Kollath}}]{Sentef2017}%
  \BibitemOpen
  \bibfield  {author} {\bibinfo {author} {\bibfnamefont {M.~A.}\ \bibnamefont
  {Sentef}}, \bibinfo {author} {\bibfnamefont {A.}~\bibnamefont {Tokuno}},
  \bibinfo {author} {\bibfnamefont {A.}~\bibnamefont {Georges}},\ and\ \bibinfo
  {author} {\bibfnamefont {C.}~\bibnamefont {Kollath}},\ }\href
  {https://doi.org/10.1103/PhysRevLett.118.087002} {\bibfield  {journal}
  {\bibinfo  {journal} {Phys. Rev. Lett.}\ }\textbf {\bibinfo {volume} {118}},\
  \bibinfo {pages} {087002} (\bibinfo {year} {2017})}\BibitemShut {NoStop}%
\bibitem [{\citenamefont {Babadi}\ \emph {et~al.}(2017)\citenamefont {Babadi},
  \citenamefont {Knap}, \citenamefont {Martin}, \citenamefont {Refael},\ and\
  \citenamefont {Demler}}]{Babadi2017}%
  \BibitemOpen
  \bibfield  {author} {\bibinfo {author} {\bibfnamefont {M.}~\bibnamefont
  {Babadi}}, \bibinfo {author} {\bibfnamefont {M.}~\bibnamefont {Knap}},
  \bibinfo {author} {\bibfnamefont {I.}~\bibnamefont {Martin}}, \bibinfo
  {author} {\bibfnamefont {G.}~\bibnamefont {Refael}},\ and\ \bibinfo {author}
  {\bibfnamefont {E.}~\bibnamefont {Demler}},\ }\href
  {https://doi.org/10.1103/PhysRevB.96.014512} {\bibfield  {journal} {\bibinfo
  {journal} {Phys. Rev. B}\ }\textbf {\bibinfo {volume} {96}},\ \bibinfo
  {pages} {014512} (\bibinfo {year} {2017})}\BibitemShut {NoStop}%
\bibitem [{\citenamefont {Murakami}\ \emph {et~al.}(2017)\citenamefont
  {Murakami}, \citenamefont {Tsuji}, \citenamefont {Eckstein},\ and\
  \citenamefont {Werner}}]{Murakami2017}%
  \BibitemOpen
  \bibfield  {author} {\bibinfo {author} {\bibfnamefont {Y.}~\bibnamefont
  {Murakami}}, \bibinfo {author} {\bibfnamefont {N.}~\bibnamefont {Tsuji}},
  \bibinfo {author} {\bibfnamefont {M.}~\bibnamefont {Eckstein}},\ and\
  \bibinfo {author} {\bibfnamefont {P.}~\bibnamefont {Werner}},\ }\href
  {https://doi.org/10.1103/PhysRevB.96.045125} {\bibfield  {journal} {\bibinfo
  {journal} {Phys. Rev. B}\ }\textbf {\bibinfo {volume} {96}},\ \bibinfo
  {pages} {045125} (\bibinfo {year} {2017})}\BibitemShut {NoStop}%
\bibitem [{\citenamefont {Curtis}\ \emph {et~al.}(2022)\citenamefont {Curtis},
  \citenamefont {Grankin}, \citenamefont {Poniatowski}, \citenamefont
  {Galitski}, \citenamefont {Narang},\ and\ \citenamefont
  {Demler}}]{Curtis2022}%
  \BibitemOpen
  \bibfield  {author} {\bibinfo {author} {\bibfnamefont {J.~B.}\ \bibnamefont
  {Curtis}}, \bibinfo {author} {\bibfnamefont {A.}~\bibnamefont {Grankin}},
  \bibinfo {author} {\bibfnamefont {N.~R.}\ \bibnamefont {Poniatowski}},
  \bibinfo {author} {\bibfnamefont {V.~M.}\ \bibnamefont {Galitski}}, \bibinfo
  {author} {\bibfnamefont {P.}~\bibnamefont {Narang}},\ and\ \bibinfo {author}
  {\bibfnamefont {E.}~\bibnamefont {Demler}},\ }\href
  {https://doi.org/10.1103/PhysRevResearch.4.013101} {\bibfield  {journal}
  {\bibinfo  {journal} {Phys. Rev. Research}\ }\textbf {\bibinfo {volume}
  {4}},\ \bibinfo {pages} {013101} (\bibinfo {year} {2022})}\BibitemShut
  {NoStop}%
\bibitem [{\citenamefont {Buzzi}\ \emph {et~al.}(2020)\citenamefont {Buzzi},
  \citenamefont {Nicoletti}, \citenamefont {Fechner}, \citenamefont
  {Tancogne-Dejean}, \citenamefont {Sentef}, \citenamefont {Georges},
  \citenamefont {Biesner}, \citenamefont {Uykur}, \citenamefont {Dressel},
  \citenamefont {Henderson}, \citenamefont {Siegrist}, \citenamefont
  {Schlueter}, \citenamefont {Miyagawa}, \citenamefont {Kanoda}, \citenamefont
  {Nam}, \citenamefont {Ardavan}, \citenamefont {Coulthard}, \citenamefont
  {Tindall}, \citenamefont {Schlawin}, \citenamefont {Jaksch},\ and\
  \citenamefont {Cavalleri}}]{Buzzi2020}%
  \BibitemOpen
  \bibfield  {author} {\bibinfo {author} {\bibfnamefont {M.}~\bibnamefont
  {Buzzi}}, \bibinfo {author} {\bibfnamefont {D.}~\bibnamefont {Nicoletti}},
  \bibinfo {author} {\bibfnamefont {M.}~\bibnamefont {Fechner}}, \bibinfo
  {author} {\bibfnamefont {N.}~\bibnamefont {Tancogne-Dejean}}, \bibinfo
  {author} {\bibfnamefont {M.~A.}\ \bibnamefont {Sentef}}, \bibinfo {author}
  {\bibfnamefont {A.}~\bibnamefont {Georges}}, \bibinfo {author} {\bibfnamefont
  {T.}~\bibnamefont {Biesner}}, \bibinfo {author} {\bibfnamefont
  {E.}~\bibnamefont {Uykur}}, \bibinfo {author} {\bibfnamefont
  {M.}~\bibnamefont {Dressel}}, \bibinfo {author} {\bibfnamefont
  {A.}~\bibnamefont {Henderson}}, \bibinfo {author} {\bibfnamefont
  {T.}~\bibnamefont {Siegrist}}, \bibinfo {author} {\bibfnamefont {J.~A.}\
  \bibnamefont {Schlueter}}, \bibinfo {author} {\bibfnamefont {K.}~\bibnamefont
  {Miyagawa}}, \bibinfo {author} {\bibfnamefont {K.}~\bibnamefont {Kanoda}},
  \bibinfo {author} {\bibfnamefont {M.-S.}\ \bibnamefont {Nam}}, \bibinfo
  {author} {\bibfnamefont {A.}~\bibnamefont {Ardavan}}, \bibinfo {author}
  {\bibfnamefont {J.}~\bibnamefont {Coulthard}}, \bibinfo {author}
  {\bibfnamefont {J.}~\bibnamefont {Tindall}}, \bibinfo {author} {\bibfnamefont
  {F.}~\bibnamefont {Schlawin}}, \bibinfo {author} {\bibfnamefont
  {D.}~\bibnamefont {Jaksch}},\ and\ \bibinfo {author} {\bibfnamefont
  {A.}~\bibnamefont {Cavalleri}},\ }\href
  {https://doi.org/10.1103/PhysRevX.10.031028} {\bibfield  {journal} {\bibinfo
  {journal} {Phys. Rev. X}\ }\textbf {\bibinfo {volume} {10}},\ \bibinfo
  {pages} {031028} (\bibinfo {year} {2020})}\BibitemShut {NoStop}%
\bibitem [{\citenamefont {Kockum}\ \emph {et~al.}(2019)\citenamefont {Kockum},
  \citenamefont {Miranowicz}, \citenamefont {Liberato}, \citenamefont
  {Savasta},\ and\ \citenamefont {Nori}}]{Kockum2019}%
  \BibitemOpen
  \bibfield  {author} {\bibinfo {author} {\bibfnamefont {A.~F.}\ \bibnamefont
  {Kockum}}, \bibinfo {author} {\bibfnamefont {A.}~\bibnamefont {Miranowicz}},
  \bibinfo {author} {\bibfnamefont {S.~D.}\ \bibnamefont {Liberato}}, \bibinfo
  {author} {\bibfnamefont {S.}~\bibnamefont {Savasta}},\ and\ \bibinfo {author}
  {\bibfnamefont {F.}~\bibnamefont {Nori}},\ }\href
  {https://doi.org/10.1038/s42254-018-0006-2} {\bibfield  {journal} {\bibinfo
  {journal} {Nat. Rev. Phys.}\ }\textbf {\bibinfo {volume} {1}},\ \bibinfo
  {pages} {19} (\bibinfo {year} {2019})}\BibitemShut {NoStop}%
\bibitem [{\citenamefont {Forn-D\'{\i}az}\ \emph {et~al.}(2019)\citenamefont
  {Forn-D\'{\i}az}, \citenamefont {Lamata}, \citenamefont {Rico}, \citenamefont
  {Kono},\ and\ \citenamefont {Solano}}]{Diaz2019}%
  \BibitemOpen
  \bibfield  {author} {\bibinfo {author} {\bibfnamefont {P.}~\bibnamefont
  {Forn-D\'{\i}az}}, \bibinfo {author} {\bibfnamefont {L.}~\bibnamefont
  {Lamata}}, \bibinfo {author} {\bibfnamefont {E.}~\bibnamefont {Rico}},
  \bibinfo {author} {\bibfnamefont {J.}~\bibnamefont {Kono}},\ and\ \bibinfo
  {author} {\bibfnamefont {E.}~\bibnamefont {Solano}},\ }\href
  {https://doi.org/10.1103/RevModPhys.91.025005} {\bibfield  {journal}
  {\bibinfo  {journal} {Rev. Mod. Phys.}\ }\textbf {\bibinfo {volume} {91}},\
  \bibinfo {pages} {025005} (\bibinfo {year} {2019})}\BibitemShut {NoStop}%
\bibitem [{\citenamefont {Thomas}\ \emph {et~al.}(2019)\citenamefont {Thomas},
  \citenamefont {Devaux}, \citenamefont {Nagarajan}, \citenamefont {Chervy},
  \citenamefont {Seidel}, \citenamefont {Hagenm{\"u}ller}, \citenamefont
  {Sch{\"u}tz}, \citenamefont {Schachenmayer}, \citenamefont {Genet},
  \citenamefont {Pupillo},\ and\ \citenamefont {Ebbesen}}]{Thomas2019}%
  \BibitemOpen
  \bibfield  {author} {\bibinfo {author} {\bibfnamefont {A.}~\bibnamefont
  {Thomas}}, \bibinfo {author} {\bibfnamefont {E.}~\bibnamefont {Devaux}},
  \bibinfo {author} {\bibfnamefont {K.}~\bibnamefont {Nagarajan}}, \bibinfo
  {author} {\bibfnamefont {T.}~\bibnamefont {Chervy}}, \bibinfo {author}
  {\bibfnamefont {M.}~\bibnamefont {Seidel}}, \bibinfo {author} {\bibfnamefont
  {D.}~\bibnamefont {Hagenm{\"u}ller}}, \bibinfo {author} {\bibfnamefont
  {S.}~\bibnamefont {Sch{\"u}tz}}, \bibinfo {author} {\bibfnamefont
  {J.}~\bibnamefont {Schachenmayer}}, \bibinfo {author} {\bibfnamefont
  {C.}~\bibnamefont {Genet}}, \bibinfo {author} {\bibfnamefont
  {G.}~\bibnamefont {Pupillo}},\ and\ \bibinfo {author} {\bibfnamefont
  {.~T.~W.}\ \bibnamefont {Ebbesen}},\ }\href@noop {} {} (\bibinfo {year}
  {2019}),\ \Eprint {https://arxiv.org/abs/1911.01459v2} {arXiv:1911.01459v2
  [cond-mat.supr-con]} \BibitemShut {NoStop}%
\bibitem [{\citenamefont {Schlawin}\ \emph {et~al.}(2019)\citenamefont
  {Schlawin}, \citenamefont {Cavalleri},\ and\ \citenamefont
  {Jaksch}}]{Schlawin2019}%
  \BibitemOpen
  \bibfield  {author} {\bibinfo {author} {\bibfnamefont {F.}~\bibnamefont
  {Schlawin}}, \bibinfo {author} {\bibfnamefont {A.}~\bibnamefont
  {Cavalleri}},\ and\ \bibinfo {author} {\bibfnamefont {D.}~\bibnamefont
  {Jaksch}},\ }\href {https://doi.org/10.1103/PhysRevLett.122.133602}
  {\bibfield  {journal} {\bibinfo  {journal} {Phys. Rev. Lett.}\ }\textbf
  {\bibinfo {volume} {122}},\ \bibinfo {pages} {133602} (\bibinfo {year}
  {2019})}\BibitemShut {NoStop}%
\bibitem [{\citenamefont {Li}\ and\ \citenamefont {Eckstein}(2020)}]{Li2020}%
  \BibitemOpen
  \bibfield  {author} {\bibinfo {author} {\bibfnamefont {J.}~\bibnamefont
  {Li}}\ and\ \bibinfo {author} {\bibfnamefont {M.}~\bibnamefont {Eckstein}},\
  }\href {https://doi.org/10.1103/PhysRevLett.125.217402} {\bibfield  {journal}
  {\bibinfo  {journal} {Phys. Rev. Lett.}\ }\textbf {\bibinfo {volume} {125}},\
  \bibinfo {pages} {217402} (\bibinfo {year} {2020})}\BibitemShut {NoStop}%
\bibitem [{\citenamefont {Curtis}\ \emph {et~al.}(2019)\citenamefont {Curtis},
  \citenamefont {Raines}, \citenamefont {Allocca}, \citenamefont {Hafezi},\
  and\ \citenamefont {Galitski}}]{Curtis2019}%
  \BibitemOpen
  \bibfield  {author} {\bibinfo {author} {\bibfnamefont {J.~B.}\ \bibnamefont
  {Curtis}}, \bibinfo {author} {\bibfnamefont {Z.~M.}\ \bibnamefont {Raines}},
  \bibinfo {author} {\bibfnamefont {A.~A.}\ \bibnamefont {Allocca}}, \bibinfo
  {author} {\bibfnamefont {M.}~\bibnamefont {Hafezi}},\ and\ \bibinfo {author}
  {\bibfnamefont {V.~M.}\ \bibnamefont {Galitski}},\ }\href
  {https://doi.org/10.1103/PhysRevLett.122.167002} {\bibfield  {journal}
  {\bibinfo  {journal} {Phys. Rev. Lett.}\ }\textbf {\bibinfo {volume} {122}},\
  \bibinfo {pages} {167002} (\bibinfo {year} {2019})}\BibitemShut {NoStop}%
\bibitem [{\citenamefont {Sentef}\ \emph {et~al.}(2020)\citenamefont {Sentef},
  \citenamefont {Li}, \citenamefont {K{\"u}nzel},\ and\ \citenamefont
  {Eckstein}}]{Sentef2020}%
  \BibitemOpen
  \bibfield  {author} {\bibinfo {author} {\bibfnamefont {M.~A.}\ \bibnamefont
  {Sentef}}, \bibinfo {author} {\bibfnamefont {J.}~\bibnamefont {Li}}, \bibinfo
  {author} {\bibfnamefont {F.}~\bibnamefont {K{\"u}nzel}},\ and\ \bibinfo
  {author} {\bibfnamefont {M.}~\bibnamefont {Eckstein}},\ }\href
  {https://doi.org/10.1103/PhysRevResearch.2.033033} {\bibfield  {journal}
  {\bibinfo  {journal} {Phys. Rev. Research}\ }\textbf {\bibinfo {volume}
  {2}},\ \bibinfo {pages} {033033} (\bibinfo {year} {2020})}\BibitemShut
  {NoStop}%
\bibitem [{\citenamefont {Mazza}\ and\ \citenamefont
  {Georges}(2019)}]{Mazza2019}%
  \BibitemOpen
  \bibfield  {author} {\bibinfo {author} {\bibfnamefont {G.}~\bibnamefont
  {Mazza}}\ and\ \bibinfo {author} {\bibfnamefont {A.}~\bibnamefont
  {Georges}},\ }\href {https://doi.org/10.1103/PhysRevLett.122.017401}
  {\bibfield  {journal} {\bibinfo  {journal} {Phys. Rev. Lett.}\ }\textbf
  {\bibinfo {volume} {122}},\ \bibinfo {pages} {017401} (\bibinfo {year}
  {2019})}\BibitemShut {NoStop}%
\bibitem [{\citenamefont {Sentef}\ \emph {et~al.}(2018)\citenamefont {Sentef},
  \citenamefont {Ruggenthaler},\ and\ \citenamefont {Rubio}}]{Sentef2018}%
  \BibitemOpen
  \bibfield  {author} {\bibinfo {author} {\bibfnamefont {M.~A.}\ \bibnamefont
  {Sentef}}, \bibinfo {author} {\bibfnamefont {M.}~\bibnamefont
  {Ruggenthaler}},\ and\ \bibinfo {author} {\bibfnamefont {A.}~\bibnamefont
  {Rubio}},\ }\href {https://doi.org/10.1126/sciadv.aau6969} {\bibfield
  {journal} {\bibinfo  {journal} {Sci. Adv.}\ }\textbf {\bibinfo {volume}
  {4}},\ \bibinfo {pages} {eaau6969} (\bibinfo {year} {2018})}\BibitemShut
  {NoStop}%
\bibitem [{\citenamefont {Schlawin}\ \emph {et~al.}(2022)\citenamefont
  {Schlawin}, \citenamefont {Kennes},\ and\ \citenamefont
  {Sentef}}]{Schlawin2022}%
  \BibitemOpen
  \bibfield  {author} {\bibinfo {author} {\bibfnamefont {F.}~\bibnamefont
  {Schlawin}}, \bibinfo {author} {\bibfnamefont {D.~M.}\ \bibnamefont
  {Kennes}},\ and\ \bibinfo {author} {\bibfnamefont {M.~A.}\ \bibnamefont
  {Sentef}},\ }\href {https://doi.org/10.1063/5.0083825} {\bibfield  {journal}
  {\bibinfo  {journal} {Appl. Phys. Rev.}\ }\textbf {\bibinfo {volume} {9}},\
  \bibinfo {pages} {011312} (\bibinfo {year} {2022})}\BibitemShut {NoStop}%
\bibitem [{\citenamefont {{Dias da Silva}}\ and\ \citenamefont
  {Dagotto}(2009)}]{Silva2009}%
  \BibitemOpen
  \bibfield  {author} {\bibinfo {author} {\bibfnamefont {L.~G. G.~V.}\
  \bibnamefont {{Dias da Silva}}}\ and\ \bibinfo {author} {\bibfnamefont
  {E.}~\bibnamefont {Dagotto}},\ }\href
  {https://doi.org/10.1103/PhysRevB.79.155302} {\bibfield  {journal} {\bibinfo
  {journal} {Phys. Rev. B}\ }\textbf {\bibinfo {volume} {79}},\ \bibinfo
  {pages} {155302} (\bibinfo {year} {2009})}\BibitemShut {NoStop}%
\bibitem [{\citenamefont {Ji}\ \emph {et~al.}(2008)\citenamefont {Ji},
  \citenamefont {Zhang}, \citenamefont {Fu}, \citenamefont {Chen},
  \citenamefont {Ma}, \citenamefont {Li}, \citenamefont {Duan}, \citenamefont
  {Jia},\ and\ \citenamefont {Xue}}]{Ji2008}%
  \BibitemOpen
  \bibfield  {author} {\bibinfo {author} {\bibfnamefont {S.-H.}\ \bibnamefont
  {Ji}}, \bibinfo {author} {\bibfnamefont {T.}~\bibnamefont {Zhang}}, \bibinfo
  {author} {\bibfnamefont {Y.-S.}\ \bibnamefont {Fu}}, \bibinfo {author}
  {\bibfnamefont {X.}~\bibnamefont {Chen}}, \bibinfo {author} {\bibfnamefont
  {X.-C.}\ \bibnamefont {Ma}}, \bibinfo {author} {\bibfnamefont
  {J.}~\bibnamefont {Li}}, \bibinfo {author} {\bibfnamefont {W.-H.}\
  \bibnamefont {Duan}}, \bibinfo {author} {\bibfnamefont {J.-F.}\ \bibnamefont
  {Jia}},\ and\ \bibinfo {author} {\bibfnamefont {Q.-K.}\ \bibnamefont {Xue}},\
  }\href {https://doi.org/10.1103/PhysRevLett.100.226801} {\bibfield  {journal}
  {\bibinfo  {journal} {Phys. Rev. Lett.}\ }\textbf {\bibinfo {volume} {100}},\
  \bibinfo {pages} {226801} (\bibinfo {year} {2008})}\BibitemShut {NoStop}%
\bibitem [{\citenamefont {Ruby}\ \emph {et~al.}(2016)\citenamefont {Ruby},
  \citenamefont {Peng}, \citenamefont {von Oppen}, \citenamefont {Heinrich},\
  and\ \citenamefont {Franke}}]{Ruby2016}%
  \BibitemOpen
  \bibfield  {author} {\bibinfo {author} {\bibfnamefont {M.}~\bibnamefont
  {Ruby}}, \bibinfo {author} {\bibfnamefont {Y.}~\bibnamefont {Peng}}, \bibinfo
  {author} {\bibfnamefont {F.}~\bibnamefont {von Oppen}}, \bibinfo {author}
  {\bibfnamefont {B.~W.}\ \bibnamefont {Heinrich}},\ and\ \bibinfo {author}
  {\bibfnamefont {K.~J.}\ \bibnamefont {Franke}},\ }\href
  {https://doi.org/10.1103/PhysRevLett.117.186801} {\bibfield  {journal}
  {\bibinfo  {journal} {Phys. Rev. Lett.}\ }\textbf {\bibinfo {volume} {117}},\
  \bibinfo {pages} {186801} (\bibinfo {year} {2016})}\BibitemShut {NoStop}%
\bibitem [{\citenamefont {Choi}\ \emph {et~al.}(2017)\citenamefont {Choi},
  \citenamefont {Rubio-Verd{\'u}}, \citenamefont {de~Bruijckere}, \citenamefont
  {Ugeda}, \citenamefont {Lorente},\ and\ \citenamefont {Pascual}}]{Choi2017}%
  \BibitemOpen
  \bibfield  {author} {\bibinfo {author} {\bibfnamefont {D.-J.}\ \bibnamefont
  {Choi}}, \bibinfo {author} {\bibfnamefont {C.}~\bibnamefont
  {Rubio-Verd{\'u}}}, \bibinfo {author} {\bibfnamefont {J.}~\bibnamefont
  {de~Bruijckere}}, \bibinfo {author} {\bibfnamefont {M.~M.}\ \bibnamefont
  {Ugeda}}, \bibinfo {author} {\bibfnamefont {N.}~\bibnamefont {Lorente}},\
  and\ \bibinfo {author} {\bibfnamefont {J.~I.}\ \bibnamefont {Pascual}},\
  }\href {https://doi.org/10.1038/ncomms15175} {\bibfield  {journal} {\bibinfo
  {journal} {Nat. Commun.}\ }\textbf {\bibinfo {volume} {8}},\ \bibinfo {pages}
  {15175} (\bibinfo {year} {2017})}\BibitemShut {NoStop}%
\bibitem [{\citenamefont {Arrachea}(2021)}]{Arrachea2021}%
  \BibitemOpen
  \bibfield  {author} {\bibinfo {author} {\bibfnamefont {L.}~\bibnamefont
  {Arrachea}},\ }\href {https://doi.org/10.1103/PhysRevB.104.134515} {\bibfield
   {journal} {\bibinfo  {journal} {Phys. Rev. B}\ }\textbf {\bibinfo {volume}
  {104}},\ \bibinfo {pages} {134515} (\bibinfo {year} {2021})}\BibitemShut
  {NoStop}%
\bibitem [{\citenamefont {Saunderson}\ \emph {et~al.}(2022)\citenamefont
  {Saunderson}, \citenamefont {Annett}, \citenamefont {Csire},\ and\
  \citenamefont {Gradhand}}]{Saunderson2022}%
  \BibitemOpen
  \bibfield  {author} {\bibinfo {author} {\bibfnamefont {T.~G.}\ \bibnamefont
  {Saunderson}}, \bibinfo {author} {\bibfnamefont {J.~F.}\ \bibnamefont
  {Annett}}, \bibinfo {author} {\bibfnamefont {G.}~\bibnamefont {Csire}},\ and\
  \bibinfo {author} {\bibfnamefont {M.}~\bibnamefont {Gradhand}},\ }\href
  {https://doi.org/10.1103/PhysRevB.105.014424} {\bibfield  {journal} {\bibinfo
   {journal} {Phys. Rev. B}\ }\textbf {\bibinfo {volume} {105}},\ \bibinfo
  {pages} {014424} (\bibinfo {year} {2022})}\BibitemShut {NoStop}%
\bibitem [{\citenamefont {Beck}\ \emph {et~al.}(2021)\citenamefont {Beck},
  \citenamefont {Schneider}, \citenamefont {R{\'o}zsa}, \citenamefont
  {Palot{\'a}s}, \citenamefont {L{\'a}szl{\'o}ffy}, \citenamefont {Szunyogh},
  \citenamefont {Wiebe},\ and\ \citenamefont {Wiesendanger}}]{Beck2021}%
  \BibitemOpen
  \bibfield  {author} {\bibinfo {author} {\bibfnamefont {P.}~\bibnamefont
  {Beck}}, \bibinfo {author} {\bibfnamefont {L.}~\bibnamefont {Schneider}},
  \bibinfo {author} {\bibfnamefont {L.}~\bibnamefont {R{\'o}zsa}}, \bibinfo
  {author} {\bibfnamefont {K.}~\bibnamefont {Palot{\'a}s}}, \bibinfo {author}
  {\bibfnamefont {A.}~\bibnamefont {L{\'a}szl{\'o}ffy}}, \bibinfo {author}
  {\bibfnamefont {L.}~\bibnamefont {Szunyogh}}, \bibinfo {author}
  {\bibfnamefont {J.}~\bibnamefont {Wiebe}},\ and\ \bibinfo {author}
  {\bibfnamefont {R.}~\bibnamefont {Wiesendanger}},\ }\href
  {https://doi.org/10.1038/s41467-021-22261-6} {\bibfield  {journal} {\bibinfo
  {journal} {Nat. Commun.}\ }\textbf {\bibinfo {volume} {12}},\ \bibinfo
  {pages} {2040} (\bibinfo {year} {2021})}\BibitemShut {NoStop}%
\bibitem [{\citenamefont {Moca}\ \emph {et~al.}(2008)\citenamefont {Moca},
  \citenamefont {Demler}, \citenamefont {Jank\'o},\ and\ \citenamefont
  {Zar\'and}}]{Moca2008}%
  \BibitemOpen
  \bibfield  {author} {\bibinfo {author} {\bibfnamefont {C.~P.}\ \bibnamefont
  {Moca}}, \bibinfo {author} {\bibfnamefont {E.}~\bibnamefont {Demler}},
  \bibinfo {author} {\bibfnamefont {B.}~\bibnamefont {Jank\'o}},\ and\ \bibinfo
  {author} {\bibfnamefont {G.}~\bibnamefont {Zar\'and}},\ }\href
  {https://doi.org/10.1103/PhysRevB.77.174516} {\bibfield  {journal} {\bibinfo
  {journal} {Phys. Rev. B}\ }\textbf {\bibinfo {volume} {77}},\ \bibinfo
  {pages} {174516} (\bibinfo {year} {2008})}\BibitemShut {NoStop}%
\bibitem [{\citenamefont {Sakurai}(1970)}]{Sakurai1970}%
  \BibitemOpen
  \bibfield  {author} {\bibinfo {author} {\bibfnamefont {A.}~\bibnamefont
  {Sakurai}},\ }\href {https://doi.org/10.1143/PTP.44.1472} {\bibfield
  {journal} {\bibinfo  {journal} {Prog. Theor. Phys.}\ }\textbf {\bibinfo
  {volume} {44}},\ \bibinfo {pages} {1472} (\bibinfo {year}
  {1970})}\BibitemShut {NoStop}%
\bibitem [{\citenamefont {Maccaferri}\ \emph {et~al.}(2021)\citenamefont
  {Maccaferri}, \citenamefont {Barbillon}, \citenamefont {Koya}, \citenamefont
  {Lu}, \citenamefont {Acuna},\ and\ \citenamefont {Garoli}}]{Maccaferri2021}%
  \BibitemOpen
  \bibfield  {author} {\bibinfo {author} {\bibfnamefont {N.}~\bibnamefont
  {Maccaferri}}, \bibinfo {author} {\bibfnamefont {G.}~\bibnamefont
  {Barbillon}}, \bibinfo {author} {\bibfnamefont {A.~N.}\ \bibnamefont {Koya}},
  \bibinfo {author} {\bibfnamefont {G.}~\bibnamefont {Lu}}, \bibinfo {author}
  {\bibfnamefont {G.~P.}\ \bibnamefont {Acuna}},\ and\ \bibinfo {author}
  {\bibfnamefont {D.}~\bibnamefont {Garoli}},\ }\href
  {https://doi.org/10.1039/D0NA00715C} {\bibfield  {journal} {\bibinfo
  {journal} {Nanoscale Adv.}\ }\textbf {\bibinfo {volume} {3}},\ \bibinfo
  {pages} {633} (\bibinfo {year} {2021})}\BibitemShut {NoStop}%
\bibitem [{\citenamefont {Benz}\ \emph {et~al.}(2016)\citenamefont {Benz},
  \citenamefont {Schmidt}, \citenamefont {Dreismann}, \citenamefont
  {Chikkaraddy}, \citenamefont {Zhang}, \citenamefont {Demetriadou},
  \citenamefont {Carnegie}, \citenamefont {Ohadi}, \citenamefont {de~Nijs},
  \citenamefont {Esteban}, \citenamefont {Aizpurua},\ and\ \citenamefont
  {Baumberg}}]{Benz2016}%
  \BibitemOpen
  \bibfield  {author} {\bibinfo {author} {\bibfnamefont {F.}~\bibnamefont
  {Benz}}, \bibinfo {author} {\bibfnamefont {M.~K.}\ \bibnamefont {Schmidt}},
  \bibinfo {author} {\bibfnamefont {A.}~\bibnamefont {Dreismann}}, \bibinfo
  {author} {\bibfnamefont {R.}~\bibnamefont {Chikkaraddy}}, \bibinfo {author}
  {\bibfnamefont {Y.}~\bibnamefont {Zhang}}, \bibinfo {author} {\bibfnamefont
  {A.}~\bibnamefont {Demetriadou}}, \bibinfo {author} {\bibfnamefont
  {C.}~\bibnamefont {Carnegie}}, \bibinfo {author} {\bibfnamefont
  {H.}~\bibnamefont {Ohadi}}, \bibinfo {author} {\bibfnamefont
  {B.}~\bibnamefont {de~Nijs}}, \bibinfo {author} {\bibfnamefont
  {R.}~\bibnamefont {Esteban}}, \bibinfo {author} {\bibfnamefont
  {J.}~\bibnamefont {Aizpurua}},\ and\ \bibinfo {author} {\bibfnamefont
  {J.~J.}\ \bibnamefont {Baumberg}},\ }\href
  {https://doi.org/10.1126/science.aah5243} {\bibfield  {journal} {\bibinfo
  {journal} {Science}\ }\textbf {\bibinfo {volume} {354}},\ \bibinfo {pages}
  {726} (\bibinfo {year} {2016})}\BibitemShut {NoStop}%
\bibitem [{\citenamefont {Lee}\ \emph {et~al.}(2019)\citenamefont {Lee},
  \citenamefont {Crampton}, \citenamefont {Tallarida},\ and\ \citenamefont
  {Apkarian}}]{Lee2019}%
  \BibitemOpen
  \bibfield  {author} {\bibinfo {author} {\bibfnamefont {J.}~\bibnamefont
  {Lee}}, \bibinfo {author} {\bibfnamefont {K.~T.}\ \bibnamefont {Crampton}},
  \bibinfo {author} {\bibfnamefont {N.}~\bibnamefont {Tallarida}},\ and\
  \bibinfo {author} {\bibfnamefont {V.~A.}\ \bibnamefont {Apkarian}},\ }\href
  {https://doi.org/10.1038/s41586-019-1059-9} {\bibfield  {journal} {\bibinfo
  {journal} {Nature}\ }\textbf {\bibinfo {volume} {568}},\ \bibinfo {pages}
  {78} (\bibinfo {year} {2019})}\BibitemShut {NoStop}%
\bibitem [{\citenamefont {Park}\ \emph {et~al.}(2019)\citenamefont {Park},
  \citenamefont {May}, \citenamefont {Leng}, \citenamefont {Wang},
  \citenamefont {Kropp}, \citenamefont {Gougousi}, \citenamefont {Pelton},\
  and\ \citenamefont {Raschke}}]{Park2019}%
  \BibitemOpen
  \bibfield  {author} {\bibinfo {author} {\bibfnamefont {K.-D.}\ \bibnamefont
  {Park}}, \bibinfo {author} {\bibfnamefont {M.~A.}\ \bibnamefont {May}},
  \bibinfo {author} {\bibfnamefont {H.}~\bibnamefont {Leng}}, \bibinfo {author}
  {\bibfnamefont {J.}~\bibnamefont {Wang}}, \bibinfo {author} {\bibfnamefont
  {J.~A.}\ \bibnamefont {Kropp}}, \bibinfo {author} {\bibfnamefont
  {T.}~\bibnamefont {Gougousi}}, \bibinfo {author} {\bibfnamefont
  {M.}~\bibnamefont {Pelton}},\ and\ \bibinfo {author} {\bibfnamefont {M.~B.}\
  \bibnamefont {Raschke}},\ }\href {https://doi.org/10.1126/sciadv.aav5931}
  {\bibfield  {journal} {\bibinfo  {journal} {Science Advances}\ }\textbf
  {\bibinfo {volume} {5}},\ \bibinfo {pages} {eaav5931} (\bibinfo {year}
  {2019})}\BibitemShut {NoStop}%
\bibitem [{\citenamefont {Shirley}(1965)}]{Shirley1965}%
  \BibitemOpen
  \bibfield  {author} {\bibinfo {author} {\bibfnamefont {J.~H.}\ \bibnamefont
  {Shirley}},\ }\href {https://doi.org/10.1103/PhysRev.138.B979} {\bibfield
  {journal} {\bibinfo  {journal} {Phys. Rev.}\ }\textbf {\bibinfo {volume}
  {138}},\ \bibinfo {pages} {B979} (\bibinfo {year} {1965})}\BibitemShut
  {NoStop}%
\bibitem [{\citenamefont {Kitamura}\ and\ \citenamefont
  {Aoki}(2016)}]{Kitamura2016}%
  \BibitemOpen
  \bibfield  {author} {\bibinfo {author} {\bibfnamefont {S.}~\bibnamefont
  {Kitamura}}\ and\ \bibinfo {author} {\bibfnamefont {H.}~\bibnamefont
  {Aoki}},\ }\href {https://doi.org/10.1103/PhysRevB.94.174503} {\bibfield
  {journal} {\bibinfo  {journal} {Phys. Rev. B}\ }\textbf {\bibinfo {volume}
  {94}},\ \bibinfo {pages} {174503} (\bibinfo {year} {2016})}\BibitemShut
  {NoStop}%
\bibitem [{\citenamefont {Canovi}\ \emph {et~al.}(2016)\citenamefont {Canovi},
  \citenamefont {Kollar},\ and\ \citenamefont {Eckstein}}]{Canovi2016}%
  \BibitemOpen
  \bibfield  {author} {\bibinfo {author} {\bibfnamefont {E.}~\bibnamefont
  {Canovi}}, \bibinfo {author} {\bibfnamefont {M.}~\bibnamefont {Kollar}},\
  and\ \bibinfo {author} {\bibfnamefont {M.}~\bibnamefont {Eckstein}},\ }\href
  {https://doi.org/10.1103/PhysRevE.93.012130} {\bibfield  {journal} {\bibinfo
  {journal} {Phys. Rev. E}\ }\textbf {\bibinfo {volume} {93}},\ \bibinfo
  {pages} {012130} (\bibinfo {year} {2016})}\BibitemShut {NoStop}%
\bibitem [{\citenamefont {Bukov}\ \emph {et~al.}(2016)\citenamefont {Bukov},
  \citenamefont {Kolodrubetz},\ and\ \citenamefont {Polkovnikov}}]{Bukov2016}%
  \BibitemOpen
  \bibfield  {author} {\bibinfo {author} {\bibfnamefont {M.}~\bibnamefont
  {Bukov}}, \bibinfo {author} {\bibfnamefont {M.}~\bibnamefont {Kolodrubetz}},\
  and\ \bibinfo {author} {\bibfnamefont {A.}~\bibnamefont {Polkovnikov}},\
  }\href {https://doi.org/10.1103/PhysRevLett.116.125301} {\bibfield  {journal}
  {\bibinfo  {journal} {Phys. Rev. Lett.}\ }\textbf {\bibinfo {volume} {116}},\
  \bibinfo {pages} {125301} (\bibinfo {year} {2016})}\BibitemShut {NoStop}%
\end{thebibliography}%

\onecolumngrid
\section{Appendix A: Derivation of the effective Hamiltonian} \label{Derivation of the effective Hamiltonian}
In the following, the derivation of the low-energy model of Hamiltonian given in Eqs.~\eqref{eq:H} to ~\eqref{eq:Hbos} in the main text is presented. We assume a high-frequency bosonic mode such that the real exchange of bosons with the solid state system is suppressed. The corresponding Hamiltonian is derived by using the Schrieffer-Wolff transformation $\hat{H}'=e^{\hat{S}} \hat{H} e^{-\hat{S}}$ with the anti-hermitian operator $\hat{S}$ where tunnelling to first order in $V_{\mathbf{k}}(g\hat{Q})$ and higher energy states of the impurity $\left(n_d=0\text{, }n_d=2\right)$ are projected out. We introduce the Hamiltonian of the unperturbed system $\hat{H}_0 \coloneqq \hat{H}_{\text{host}} + \hat{H}_d+ \hat{H}_{\omega_0}$ and the projection operator $\hat{P}_{n_{d},m}$ onto the subspace in which the impurity is occupied by the number of $n_d$ electrons $\left(n_d \in \lbrace 0,1,2 \rbrace\right)$ and in which the bosonic system is in the number state $\ket{m}$ with $m \geq 0$. By performing a Taylor expansion where only terms up to second order in $V_{\mathbf{k}}(g\hat{Q})$ are considered and by imposing the condition $\hat{P}_{n_{d},m}\left[ \hat{H}_{0},\hat{S}\right]\hat{P}_{n_{d}',m'}=\hat{P}_{n_{d},m}\hat{H}_{\text{hyb}}\hat{P}_{n_{d}',m'}$ where either $n_d=1$ or $n_d'=1$ holds, we obtain the effective Hamiltonian $\hat{H}_{\text{eff}}$. Expanded in the boson number basis $\left\{ \ket{m}\right\}$, it can be written as
\begin{align}
\hat{H}_{\text{eff}}=\sum_{mm'}\left(\hat{P}_{1,m}\hat{H}_{0}\hat{P}_{1,m'}+\frac{1}{2}\hat{P}_{1,m}\left[\hat{S},\hat{H}_{\text{hyb}}\right]\hat{P}_{1,m'}\right).\label{eq:Heff}
\end{align}

In the high-frequency limit, the exchange of bosons with the solid occurs only as intermediate states and we can project the Hamiltonian onto the sector of a fixed boson number $n\geq 0$. The Hamiltonian reads
\begin{align}
\hat{H}^{n,n}_{\text{eff}}\ket{n}\bra{n}\coloneqq \bra{n}\hat{H}_{\text{eff}}\ket{n}\ket{n}\bra{n}=\hat{P}_{1,n}\hat{H}_{0}\hat{P}_{1,n}+\frac{1}{2}\hat{P}_{1,n}\left[\hat{S},\hat{H}_{\text{hyb}}\right]\hat{P}_{1,n}.\label{eq:Heffnn}
\end{align}
The generator $\hat{S}$ can be determined by
\begin{align}
\bra{\underline{n}_e,\underline{n}_{d},m}\hat{P}_{n_{d},m}\hat{S}\hat{P}_{n_{d}',m'}\ket{\underline{n}_e',\underline{n}'_{d},m'}=\frac{1}{E_{\underline{n}_e,\underline{n}_{d},m}-E_{\underline{n}_e',\underline{n}_{d}',m'}}\bra{\underline{n}_e,\underline{n}_{d},m}\hat{P}_{n_{d},m}\hat{H}_{\text{hyb}}\hat{P}_{n_{d}',m'}\ket{\underline{n}_e',\underline{n}'_{d},m'}.\label{eq:S}
\end{align}
We introduced the eigenket $\ket{\underline{n}_e,\underline{n}_{d},m}$ with the eigenenergy $E_{\underline{n}_e,\underline{n}_{d},m}$ to the unperturbed system $\hat{H}_0$, where $\underline{n}_e$ describes excitations in the substrate, $\underline{n}_{d}$ is the occupation of the impurity and $m$ corresponds to the number of bosons. In order to calculate Eq.~\eqref{eq:S}, we expand the Hamiltonian in the boson number basis,
\begin{align}
\hat{H} & = \sum_{m=0}^{\infty}\sum_{m'=0}^{\infty}\hat{H}^{m,m'}\ket{m}\bra{m'},\\
\hat{H}^{m,m'} & = \hat{H}_{0}^{m,m'}+\hat{H}_{\text{hyb}}^{m,m'}.
\end{align}
By defining $V_{\mathbf{k},\gamma}^{m,m'} \coloneqq \bra{m}V_{\mathbf{k},\gamma}(g\hat{Q})\ket{m'}$, the anti-hermitian generator written in terms of Bogoliubov quasiparticle operators is given by
\begin{align}
\hat{S} = & \sqrt{2} \sum_{\mathbf{k}\sigma \gamma mm'}\sum_{\eta=+,-} \left\lbrace V_{\mathbf{k},\gamma}^{m,m'}\left( \frac{u_\mathbf{k}^{\ast}n_{d-\sigma}^{\eta}\hat{\alpha}_{\mathbf{k}\sigma,\gamma}^{\dagger}\hat{c}_{d\sigma}}{E_{\mathbf{k}}-\epsilon_{d}^{\eta} + \left(m-m'\right)\omega_{0}} + \frac{\sigma v_\mathbf{k}n_{d\sigma}^{\eta}\hat{\alpha}_{-\mathbf{k}-\sigma,\gamma}\hat{c}_{d\sigma}}{E_{\mathbf{k}}+\epsilon_{d}^{\eta} - \left(m-m'\right)\omega_{0}}   \right)\ket{m}\bra{m'}- h.c.\right\rbrace,
\end{align}
where we have introduced the definitions $\eta=+,-$, $n_{d\sigma}^+=n_{d\sigma}$, $n_{d\sigma}^-=1-n_{d\sigma}$, $\epsilon_{d\sigma}^+=-\epsilon_{d\sigma}$ and $\epsilon_{d\sigma}^-=\epsilon_{d\sigma}$.

Neglecting constant energy terms and considering inversion-symmetry around the impurity, we obtain two decoupled channels,
\begin{align}
\hat{H}_{0} = & \sum_{\mathbf{k}\sigma\gamma}\left(\epsilon_{\mathbf{k}}-\mu\right)\hat{a}_{\mathbf{k}\sigma,\gamma}^{\dagger}\hat{a}_{\mathbf{k}\sigma,\gamma}-\sum_{\mathbf{k},\gamma}\left(\Delta^{\ast}\hat{a}_{-\mathbf{k}\downarrow,\gamma}\hat{a}_{\mathbf{k}\uparrow,\gamma}+h.c.\right),\label{eq:H0}\\
\hat{H}_{\text{scat}}^{n,n} = & \sum_{\mathbf{k}\mathbf{k}'\sigma \gamma }\sum_{l=-n}^{\infty}\frac{1}{\epsilon_d -l\omega_0} \left[V_{\mathbf{k},\gamma}^{n,n+l}\left(V_{\mathbf{k}',\gamma}^{n,n+l}\right)^{\ast} - V_{\mathbf{k},\gamma}^{n+l,n}\left(V_{\mathbf{k}',\gamma}^{n+l,n}\right)^{\ast}\right] \hat{a}_{\mathbf{k}\sigma,\gamma}^{\dagger}\hat{a}_{\mathbf{k}'\sigma,\gamma},\\
\hat{H}_{\text{spin}}^{n,n} = & -4\sum_{\mathbf{k}\mathbf{k}'\gamma}\sum_{l=-n}^{\infty}\frac{1}{\epsilon_d -l\omega_0}\left[V_{\mathbf{k},\gamma}^{n,n+l}\left(V_{\mathbf{k}',\gamma}^{n,n+l}\right)^{\ast} + V_{\mathbf{k},\gamma}^{n+l,n}\left(V_{\mathbf{k}',\gamma}^{n+l,n}\right)^{\ast}\right]\mathbf{\hat{S}}^{\mathbf{k}\mathbf{k}'}_{\gamma}\cdot\mathbf{\hat{S}}_d,
\end{align}
with $\mathbf{\hat{S}}^{\mathbf{k}\mathbf{k}'}_{\gamma}=\sum_{\sigma\sigma'}\frac{1}{2}\hat{a}_{\mathbf{k}\sigma,\gamma}^{\dagger}\bm{\sigma}_{\sigma\sigma'}\hat{a}_{\mathbf{k'}\sigma',\gamma}$ and the even and odd parity combinations $\hat{a}_{\mathbf{k}\sigma,\gamma}$ given in the main text. We consider an impurity level $\epsilon_d$ well below the Fermi surface and a small superconducting gap (compared to $\epsilon_d$). Therefore, we used $E_{\mathbf{k}} \pm \epsilon_d \approx \pm \epsilon_d$ in the derivation. In addition, we have used that $V_{\mathbf{k},\gamma}^{n,n+l}\left(V_{\mathbf{k}',-\gamma}^{n,n+l}\right)^{\ast}=0$ holds, since $V_{\mathbf{k},+\left(-\right)}\left(g\hat{Q}\right)$ contains only even (odd) powers of $\hat{Q}$ $\left(V_{\mathbf{k},+\left(-\right)}^{n,n+l}=0\text{ for }l=\text{odd (even)}\right)$.

Taking in addition the relation $V_{\mathbf{k},\gamma}^{n,n+l}\left(V_{\mathbf{k}',\gamma}^{n,n+l}\right)^{\ast} = V_{\mathbf{k},\gamma}^{n+l,n}\left(V_{\mathbf{k}',\gamma}^{n+l,n}\right)^{\ast}$
into account, yields
\begin{align}
\hat{H}_{\text{scat}}^{n,n} = & 0,\label{eq:Hscatnnf}\\
\hat{H}_{\text{spin}}^{n,n} = & -\sum_{\mathbf{k}\mathbf{k}'\gamma}J_{n,\gamma}\mathbf{\hat{S}}^{\mathbf{k}\mathbf{k}'}_{\gamma}\cdot\mathbf{\hat{S}}_d.\label{eq:Hspinnnf}
\end{align}
The coupling constant is given by
\begin{align}
J_{n,+\left(-\right)}=\sum_{l=\text{even(odd)}}\frac{8}{\epsilon_d -l\omega_0}V_{\mathbf{k},+\left(-\right)}^{n,n+l}\left(V_{\mathbf{k}',+\left(-\right)}^{n,n+l}\right)^{\ast}.\label{eq:Jnnf}
\end{align}

In particular, the Peierls coupling given in Eq.~\eqref{eq:Peierls} in the main text yields 
\begin{align}
J_{n,+\left(-\right)} = \sum_{l=\text{even}\left(\text{odd}\right)} \frac{8{\lvert V \rvert}^2}{\epsilon_d} \frac{j_{n,n+l}^2}{1-\frac{l\omega_{0}}{\epsilon_{d}}}=\sum_{l=\text{even}\left(\text{odd}\right)} J \frac{j_{n,n+l}^2}{1-\frac{l\omega_{0}}{\epsilon_{d}}},\label{eq:JPeierls} 
\end{align}
where we defined $j_{n,n+l}$ \cite{Li2020} 
\begin{align}
j_{n,n+l} & = e^{-g^{2}/2}\sum_{s=0}^{n+l}\frac{\left(-1\right)^{s}g^{2s+\lvert l \rvert}}{s!\left(s+\lvert l \rvert\right)!}\sqrt{\frac{n!}{\left(n+l\right)!}}\frac{\left(n+l\right)!}{\left(n+l-s\right)!}\text{ for } -n\leq l \leq 0,\\
j_{n,n+l} & = e^{-g^{2}/2}\sum_{s=0}^{n}\frac{\left(-1\right)^{s}g^{2s+\lvert l \rvert}}{s!\left(s+\lvert l \rvert\right)!}\sqrt{\frac{\left(n+l\right)!}{n!}}\frac{n!}{\left(n-s\right)!}\text{ for }l>0.\label{eq:j}
\end{align}
\section{Appendix B: Classical Floquet Limit}
\begin{figure}[h]
\centering
\includegraphics[width=0.9\columnwidth]{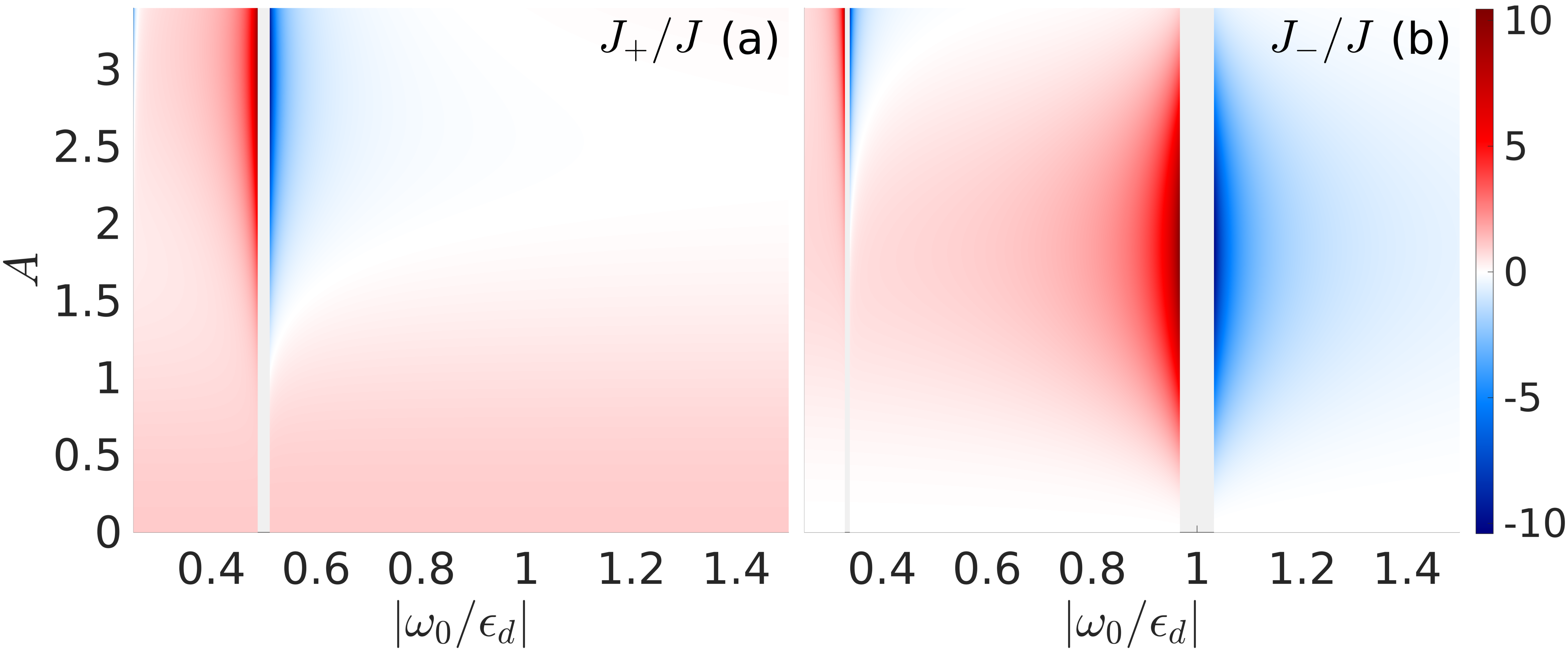}
\caption{Exchange coupling constant for the even (a) and odd (b) channel normalised by $J<0$ as a function of $A$ and $\left| \omega_{0} / \epsilon_{d} \right|$ in the classical limit. Antiferromagnetic (ferromagnetic) exchange scattering is coloured in red (blue). The undriven case corresponds to $J_{+}=1$ (red) and $J_{-}=0$ (white). The areas around the resonances ($\left| \omega_{0} / \epsilon_{d} \right|=1/2$, $\left| \omega_{0} / \epsilon_{d} \right|=1$, $\left| \omega_{0} / \epsilon_{d} \right|=1/3$) are excluded.}
	\label{CCFL}
\end{figure}

For the classically driven system, $g\hat{Q}$ is replaced by a time-dependent field $A\cos\left(\omega_{0}t\right)$, with dimensionless amplitude $A$. This gives rise to a time-periodic Hamiltonian $\hat{H}\left(t\right)=\hat{H}\left(t+T\right)$ with period $T=2\pi/\omega_0$. The stroboscopic dynamics is described by the time-independent Floquet Hamiltonian. Like in the undriven case, this Floquet Hamiltonian can be projected onto a given low energy subspace (the singly occupied impurity) using a time-periodic Schrieffer-Wolff transformation \cite{Eckstein2017,Kitamura2016,Canovi2016,Bukov2016,Bukov2015}. Alternatively, the Floquet Hamiltonian can be obtained within the Floquet block matrix approach, in which the periodically driven system is mapped to a time-independent Hamiltonian in an extended Hilbert space with an additional discrete Floquet band index. In the high-frequency limit, we can then obtain the Floquet Hamiltonian by projecting to the lowest Floquet band \cite{Bukov2015,Bukov2016}. This Floquet Hamiltonian for the Peierls coupling is reproduced from the quantum Floquet formalism by taking the limit $n\rightarrow\infty \text{, }g\rightarrow 0$ with $g\sqrt{n}$ being a finite number, which was already observed in \cite{Sentef2020,Li2020}. In this limit, $j_{n,n+l}$ converges to the $\left|l\right|$th Bessel function of the first kind $\mathcal{J}_{\left|l\right|}\left(2g\sqrt{n}\right)$,
\begin{align}
\mathcal{J}_{\left|l\right|}\left(A\right)=\sum_{k=0}^{\infty}\frac{\left(-1\right)^{k}}{k!\left(k+\left|l\right|\right)!}\left(\frac{A}{2}\right)^{2k+\left|l\right|}.
\end{align}
We obtain the effective Hamiltonian given in Eqs. (\ref{eq:H0}), (\ref{eq:Hscatnnf}) and (\ref{eq:Hspinnnf}) in App. A but with the modified coupling constant
\begin{align}
J_{+\left(-\right)} = \sum_{l=\text{even}\left(\text{odd}\right)} J \frac{\left(\mathcal{J}_{\left|l\right|}\left(A\right)\right)^{2}}{1-\frac{l\omega_{0}}{\epsilon_{d}}}.\label{eq:JPeierls_FL}
\end{align}
Figure~\ref{CCFL} depicts the exchange coupling constant normalised by the bare antiferromagnetic exchange constant $J$ as a function of $A$ and $\lvert\omega_0/\epsilon_d\rvert$ for the even (a) and odd (b) channel. The obtained results are very similar to the exchange constants for two photons in the cavity (see Fig. \ref{fig:J} in the main text). However, a resonance at $\lvert \omega_0/\epsilon_d \rvert=1/3$ is shown in the odd channel due to the infinite number of available photons. In the quantum formalism, this resonance appears for at least three photons in the cavity. These statements are transferred to the expressions for the energies.

\begin{figure}[tbp]
\centering
\includegraphics[width=0.9\columnwidth]{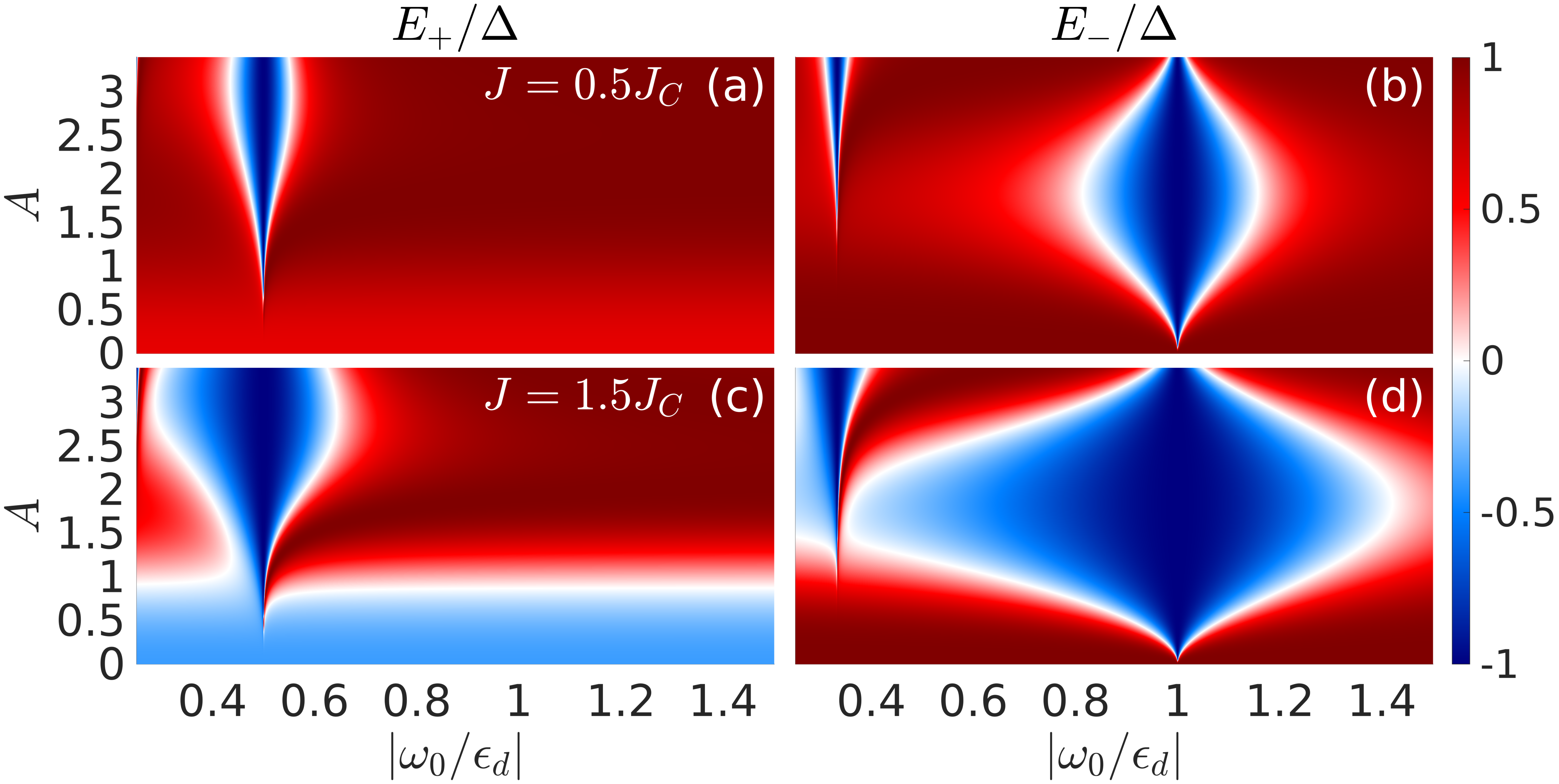}
\caption{Lowest Floquet band quasienergy for the even $E_{+}$ and odd $E_{-}$ channel normalised by $\Delta$ for $J=0.5 J_C $ (a)-(b) and $J=1.5 J_C $ (c)-(d) as a function of $A$ and $\left| \omega_{0} / \epsilon_{d} \right|$ in the classical limit. The Fermi energy $\mu$ is set to zero.}
\label{E_FL}
\end{figure}
\begin{figure}[tbp]
\centering
\includegraphics[width=0.9\columnwidth]{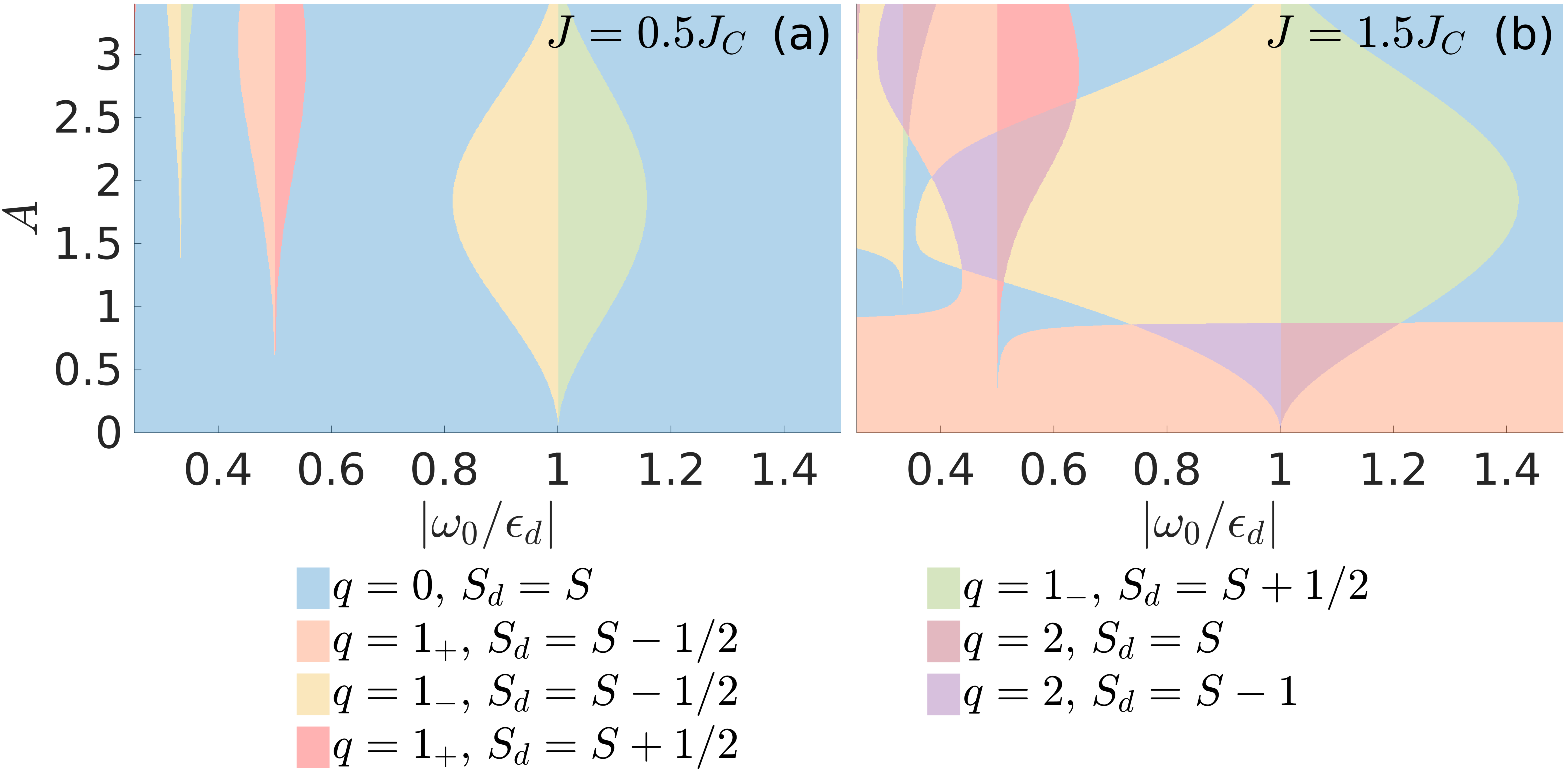}
\caption{Floquet state diagram for $J/J_C=0.5$ (a) and $J/J_C=1.5$ (b) as a function of $A$ and $\left| \omega_{0} / \epsilon_{d} \right|$ in the classical limit. $q$ denotes the number of bound quasiparticles within each channel and $S_d$ the impurity spin.}
\label{GS_FL}
\end{figure}

\end{document}